\documentclass[useAMS, usenatbib]{mn2e}
\usepackage{xspace}
\usepackage{amsmath}
\usepackage{epsfig}
\usepackage{multicol}

\usepackage{hyperref,cleveref}
\usepackage{tabularx}
\usepackage{cleveref}
\crefname{section}{\S}{\S\S}
\Crefname{section}{\S}{\S\S}

\newcommand{\ltsima}{\mbox{$\; \buildrel < \over \sim \;$}}
\def\simlt{\lower.5ex\hbox{\ltsima}}
\def\gtsima{$\; \buildrel > \over \sim \;$}
\def\simgt{\lower.5ex\hbox{\gtsima}}

\def\fesc{$\langle f_{esc}\rangle$\xspace}
\def\eq{Equation}

\def\eq{equation}

\def\eg{{\it e.g.}}
\def\ltsima{$\; \buildrel < \over \sim \;$}
\def\simlt{\lower.5ex\hbox{\ltsima}}
\def\gtsima{$\; \buildrel > \over \sim \;$}
\def\simgt{\lower.5ex\hbox{\gtsima}}
\def\eq{equation}
\def\H2{H$_2$\xspace}

\def\fesc{{$\langle f_{esc}\rangle$}\xspace}
\def\h2{H$_2$\xspace}
\def\ion#1#2{\text{#1\,\sc #2}}
\def\HI{{\ion{H}{i} }}
\def\HII{{\ion{H}{ii} }}
\def\GI{{\ion{He}{i} }}
\def\GII{{\ion{He}{ii} }}

\def\pop3{Population~III\xspace}

\def\hide#1{}

\pdfoutput=1

\title{Modeling Reionization in a Bursty Universe}

\author[Hartley \& Ricotti]{Blake Hartley$^{1}$ and Massimo Ricotti$^{1}$\\
$^1$Department of Astronomy, University of Maryland, College Park, MD 20742, USA
\thanks{bth@astro.umd.edu, ricotti@astro.umd.edu}
}
\begin{document}
\maketitle
\begin{abstract}
We present semi-analytic models of the epoch of reionization focusing on the differences between continuous and bursty star formation  (SF). Our model utilizes physically motivated analytic fits to 1D radiative transfer simulations of \HII regions around dark matter halos in a representative cosmic volume. Constraining our simulations with observed and extrapolated UV luminosity functions of high redshift galaxies, we find that for a fixed halo mass, stellar populations forming in bursty models produce larger \HII regions which leave behind long-lived relic \HII regions which are able to maintain partial ionization in the intergalactic medium (IGM) in a manner similar to an early X-ray background. The overall effect is a significant increase in the optical depth of the IGM, $\tau_e$, and a milder increase of the redshift of reionization. To produce $\tau_e =0.066$ observed by Planck and complete reionization by redshift $z_{\rm re}\sim 6$, models with bursty SF require an escape fraction \fesc $\sim 2\%-10\%$ that is 2-10 times lower than \fesc$\sim 17\%$ found assuming continuous SF and is consistent with upper limits on \fesc from observations at $z=0$ and $z\sim 1.3-6$. The ionizing photon budget needed to reproduce the observed $\tau_e$ and $z_{\rm re}$ depends on the period and duty cycle of the bursts of SF and the temperature of the neutral IGM. These results suggest that the tension between observed and predicted ionizing photon budget for reionization can be alleviated if reionization is driven by short bursts of SF, perhaps relating to the formation of Population~III stars and compact star clusters such as proto-globular clusters.
\end{abstract}
\begin{keywords}
cosmology: theory -- cosmology: early Universe -- cosmology: dark ages, reionization, first stars
\end{keywords}

\section{Introduction}\label{sec:intro}

Reionization is one of the least understood epochs of the history of the universe, mainly because the dominant sources of ionizing radiation during this epoch are currently unknown. After recombination, the density perturbations throughout the universe grew until the first stars and galaxies formed. A fraction of the UV photons produced by these objects escaped from their host halos and began ionizing the inter-galactic medium (IGM). Some combination of stars, galaxies, and quasi-stellar objects (QSOs) produced enough ionizing photons to fully ionize the IGM by a redshift of $z\sim 6$, as required by a variety of observations \citep{Fan:2002, Fan:2006, Becker:2007, Bolton:2011, Mortlock:2011,Planck:2015}. The details of this phase transition are poorly understood, as few direct observations of the universe at high redshifts are possible.
Observations of the Lyman-$\alpha$ forest absorption lines in the spectrum of distant quasars have shown unambiguously that the IGM was ionized to a very high level by a redshift not much later than $z\sim 6$ \citep[\eg,][]{Fan:2006}. Observations of Lyman-$\alpha$ emissions from sources at redshifts $z>6$ indicate rapid changes in their abundance at these redshifts \citep{Ouchi:2010, Kashikawa:2011, Caruana:2014}. These observations, as well as the optical depth to Thomson scattering on free electrons of the IGM, $\tau_e \sim 0.066$, derived from CMB observations \citep{Planck:2015} and observations of UV light from redshift $z \sim 10$ galaxies \citep{McLure:2010, Pentericci:2011, Bouwens:2013, Oesch:2014}, suggest that reionization was well underway significantly earlier. The duration of reionization and the nature of the sources which supplied the necessary photons remains the subject of observational and theoretical research \citep{Oh:00, Venkatesan:01, HansenH:03, MadauR:03, Sokasian:03, RicottiO:2004, RicottiOb:2004, RicottiO:2005, RicottiOM:2008, VolonteriG:09, Trenti:2010, Wise:2014, MBK:2015}.
There is a vast body of theoretical and modeling work on this topic, including semi-analytic models of reionization processes \citep[][]{Furlanetto:2004, 2007MNRAS.380L..30A, 2011MNRAS.410..775R, 2011MNRAS.411..955M, 2011MNRAS.414..727Z, 2013ApJ...776...81B}, a variety of different radiative transfer simulations often limited by being over static density fields \citep[][]{2001NewA....6..359S, 2003MNRAS.344L...7C, 2006MNRAS.369.1625I, 2007ApJ...654...12Z, 2007MNRAS.377.1043M, 2008MNRAS.388.1501C, 2008ApJ...689L..81T, 2010ApJ...724..244A, 2012ApJ...756L..16A}, and more recently full radiation hydrodynamics simulations which follow matter and radiation simultaneously and self consistently \citep[][]{Gnedin:2000, 2011MNRAS.415.3731P, 2013MNRAS.429L..94P, 2014ApJ...789..149S, Gnedin:2014, 2014ApJ...793...30G, 2015ApJS..216...16N, 2015MNRAS.451.1586P}. However, all these simulations treat the ionizing radiation photons as monochromatic: the 3D transfer of radiation is done on a low number (often one) of frequency bands. 

In this paper we employ a simple model to explore the possibility that the sources of reionization had an intermittent UV emissivity. Sources that are definitively characterized by a bursty mode of star formation include Population~III stars and the first small-mass dwarf galaxies \citep{RicottiGS:2002a, RicottiGS:2002b, Schaerer:2003, Oshea:2015}. Our model also includes radiation transfer of ionizing radiation well sampled in the frequency domain (about 400 logarithmically spaced frequency bins), allowing us to properly reproduce the width of ionization fronts and specific intensity of ionizing background radiation, that are affected by high energy photons.
At low redshift, the halo matching technique has proven to be a good ansatz to match observed galaxies to dark matter halos from simulations \citep{ValeO:2004,Vale0:2006, Guo:2010, Moster:2010}. This method works by successively placing the brightest stellar populations within the most massive halos, neglecting the possibility that some halos may become significantly brighter for a brief period of time. However, at high redshift the high merger rate and the small masses of the first galaxies suggest that star formation in galaxies should be rather bursty, as confirmed by simulations. In particular, it has been suggested \citep{Ricotti:2002, KatzR:2013, KatzR:2014} that the formation of compact stellar systems before reionization, which may lead relics such as globular clusters, ultra-compact dwarfs and dwarf-globular transition objects, may dominate reionization. In these scenarios we expect an effective duty cycle for UV luminosity of the first galaxies, leading to a large fraction of halos of any given mass to be nearly dark in between short lived bursts.

The simplest semi-analytic method typically used in literature to investigate the ionization evolution of the IGM is, essentially, to keep a budget of hydrogen ionizing photons needed to maintain a fraction of the volume in IGM, $Q_{\text{HII}}(t)$, fully ionized at a time $t$ \citep{Madau:1999, Kuhlen:2012}. This filling fraction evolves according to a simple differential equation:
\begin{align}\label{eq:dqdt}
\frac{dQ_{\text{HII}}}{dt}=\frac{\dot{n}_{\text{ion}}}{\bar{n}_{\text{H}}}-\frac{Q_{\text{HII}}}{\bar{t}_{\text{rec}}}.
\end{align}
Here, $\dot{n}_{\rm ion}$ is the rate of ionizing photon production per comoving volume, $\bar{n}_{\rm H}$ is the mean comoving cosmic number density of atomic hydrogen, and 
\begin{align}\label{eq:trec} 
\bar{t}_{\rm rec}&=\frac{1}{C_{\HII}\alpha_{\rm B}(T_0)\bar{n}_H(1+Y/4X)(1+z)^3}\\
&\approx0.93\,{\rm Gyr}\,\left(\frac{C_{\HII}}{3}\right)^{-1}
\left(\frac{T_0}{2\times10^4\,{\rm K}}\right)^{0.7}
\left(\frac{1+z}{7}\right)^{-3}\nonumber
\end{align}
is a time scale of hydrogen recombination in fully ionized bubbles of \HII. Here, $\alpha_{\rm B}$ is the case-B hydrogen recombination coefficient, $T_{0}$ is the IGM temperature at mean density, $C_{\HII}$ is the effective clumping factor in ionized gas, $X=0.75$ is the hydrogen mass fraction and $Y=0.25$ the helium mass fraction. We assume that helium is singly ionized at the same time as hydrogen, but only fully ionized at $z<4$ (see \S~\ref{sec:taue}).
However, the rate of hydrogen recombination is proportional to $x_e^2$, where $x_e$ is the electron fraction in the IGM (so that $n_e=x_e \bar{n}_{\rm H}$). Thus, after a burst of star formation recombination proceeds quickly at first, but slows down as lower levels of partial ionization are reached. The method described by Equation~\eqref{eq:dqdt} lumps the complicated spatial dependence of electron fraction and recombination rate  together into $\bar{t}_{\rm rec}$, and as such does not take into account the volume filling fraction of partially ionized gas and the reduced recombination rates in regions of partial ionization produced by intermittent star formation.

Here, we focus on investigating the observational implications of assuming either a continuous or a bursty mode of star formation (SF) in the first galaxies, and how each affects the evolution of the IGM during the epoch of reionization (EOR). A sudden burst of star formation produces a much higher luminosity of ionizing photons than the same mass of stars forming continuously over a period significantly longer than the lifetime of the brightest stars ($\sim 5-10$~Myr). This increased luminosity produces a larger \HII region than in the continuous case. Once the brightest stars have died, however, the larger region of ionization begins to recombine. The rate of recombination is proportional to the electron fraction, so that the electron fraction within these regions remains non-trivially boosted for long periods of time.  We wish to investigate whether these relatively long lived relic \HII regions of partially ionized gas have a signature similar to hypothetical X-ray preheating of the IGM \citep{Venkatesan:01, RicottiO:2004, RicottiO:2005} and if they produce an observable effect on the optical depth to Thompson scattering of the IGM $\tau_e$. 
The model simulates a population of dark matter halos hosting star forming populations in a representative cosmic volume between the redshifts of $z=30$ and $z\approx5.8$, a period of $900$ Myr. Luminosities are assigned through a halo matching process such that the galaxies UV luminosity functions (and the mean ionizing emissivity) in both cases are identical and match observations. In the Appendix we present physical models of cosmological \HII regions and the evolution of the relic \HII regions in the presence of bursty star formation. These physical models, calibrated to reproduce 1D radiation transfer simulations, provide useful equations for the evolution of the volume filling fraction of partially ionized gas. We use the results of these simulations to calculate the average electron fraction at a given time and the optical depth to Thompson scattering of the IGM and analyze how different modes of star formation affect these observable quantities. Near the redshift of reionization the overlap of individual \HII regions produces ionized bubbles containing many UV sources. Hence, in this final phase the time-average UV emissivity within bubbles approaches the continuous limit even if the individual galactic sources are bursty. However, this is does not affect our main results as we find that the ionization history in bursty models differs from the continuous models mostly at high-redshift, when the \HII regions around individual sources hardly overlap.

The paper is organized as follows. In \cref{sec:simulation} we describe the methods used to simulate a cosmic volume during the EOR and how observable quantities are derived from the results. Most of the technical aspects of the method and a physically motivated analytic model describing recombining \HII regions are presented in the Appendix. In \cref{sec:results}, we present results of a fiducial set of simulations and a parameters study to asses how the results depend on the free parameters in the model. In \cref{sec:summary}, we present a summary and concluding remarks. 
Throughout this paper we assume a flat $\Lambda$CDM cosmology with $\tau_e=0.066 \pm 0.012$, $H_0=67.51\text{ km s}^{-1}\text{ Mpc}^{-1}$, ($\Omega_{\Lambda}, \Omega_{m}, \Omega_{b}, n_s, \sigma_8)$=$(0.691, 0.309, 0.0489, 0.9667, 0.816)$,
as presented by the Planck Collaboration \citep{Planck:2015}. A fiducial value for the redshift of reionization $z_{\rm re}=6.0$ will be assumed as a constrain for parameter studies.

\section{Simulating a Cosmic Volume}\label{sec:simulation}

\subsection{Overview of the Methodology}

Our simulation is a numerical representation of the evolution of a cosmic volume between the redshifts of $z\sim30$ and $z\sim 5.8$, or a time of roughly 1~Gyr. The volume is a cube with side length $100$~comoving ${\rm Mpc}$ (cMpc), which is enough to expect convergence of our results \citep[\eg,][]{Iliev:2014}. Dark matter halos, extracted from a Press-Schechter distribution, are placed in the volume randomly (neglecting clustering). Newly virialized halos are added as the Hubble time increases and star formation begins with the addition of the halo. In the continuous case, star formation proceeds continuously throughout the life of the halo, whereas in the bursty case star formation occurs periodically with a given period $\Delta T$ and duty cycle $f_{\rm duty}$. In both modes of SF the time averaged ionizing emissivity of the halos is the same.

\subsection{Halo matching}

We assign a luminosity to any given halo within our cosmic volume according to a halo matching procedure \citep[see,][]{Guo:2010}. We randomly generate halos of mass $M$ at redshift $z$ with a mass distribution $\phi_M(M,z)$ using the Press-Schechter formalism with the modification by Sheth-Tormen \citep{PressS:1974, ShethT:2002} using Planck 2015 cosmological parameters. We limit the halos to a minimum mass $m_{\rm dm}$ (which we will vary to understand which halos contribute the most to our results.)

We assign a luminosity to each halo so that the population has a luminosity function described by  a Schechter function $\phi_L(L,z)$ consistent with HST deep field observations. We use fits to the Schechter parameters evolution as a function of $z$ from \citet{Kuhlen:2012}. As in their study, we consider three models: "FIT," "MAX, and "MIN." These models are extrapolations of Schechter function parameters fit to published luminosity functions at different redshifts in the rest frame UV band at 1500~\AA\ presented in \citet{Bouwens:2015}. The "FIT" model is the best linear regression for the time dependence of the Schechter parameters of the form $\{M^*,\log_{10}\phi^*,\alpha\}=A+B(z-6)$, while the "MAX" and "MIN" models independently adjust these parameters by $\pm 1\sigma$ (see \citet{Kuhlen:2012} for a discussion).

\begin{figure}
\includegraphics[width=8.0cm]{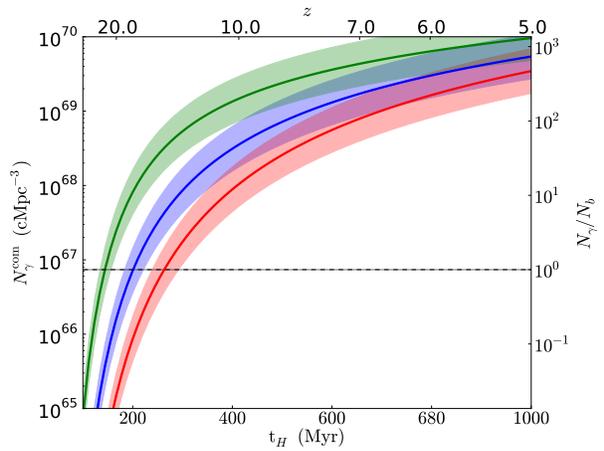}
\caption{Cumulative photon count per comoving cubic Mpc for the FIT (blue), MIN (red), and MAX (green) models presented in \citep{Kuhlen:2012}. The shaded regions represent the full range of emission hardness ($\zeta=0.5$ to $\zeta=2.0$; see text). The dotted line and accompanying shaded region shows the comoving number density of baryons as derived from the Planck results \citep{Planck:2015}. We interpret the ratio of photon count to baryon count (plotted on the right axis) as the upper limit of $1/f_{\rm esc}$.}
\label{fig:emiss}
\end{figure}

\begin{figure*}
\includegraphics[width=8.7cm]{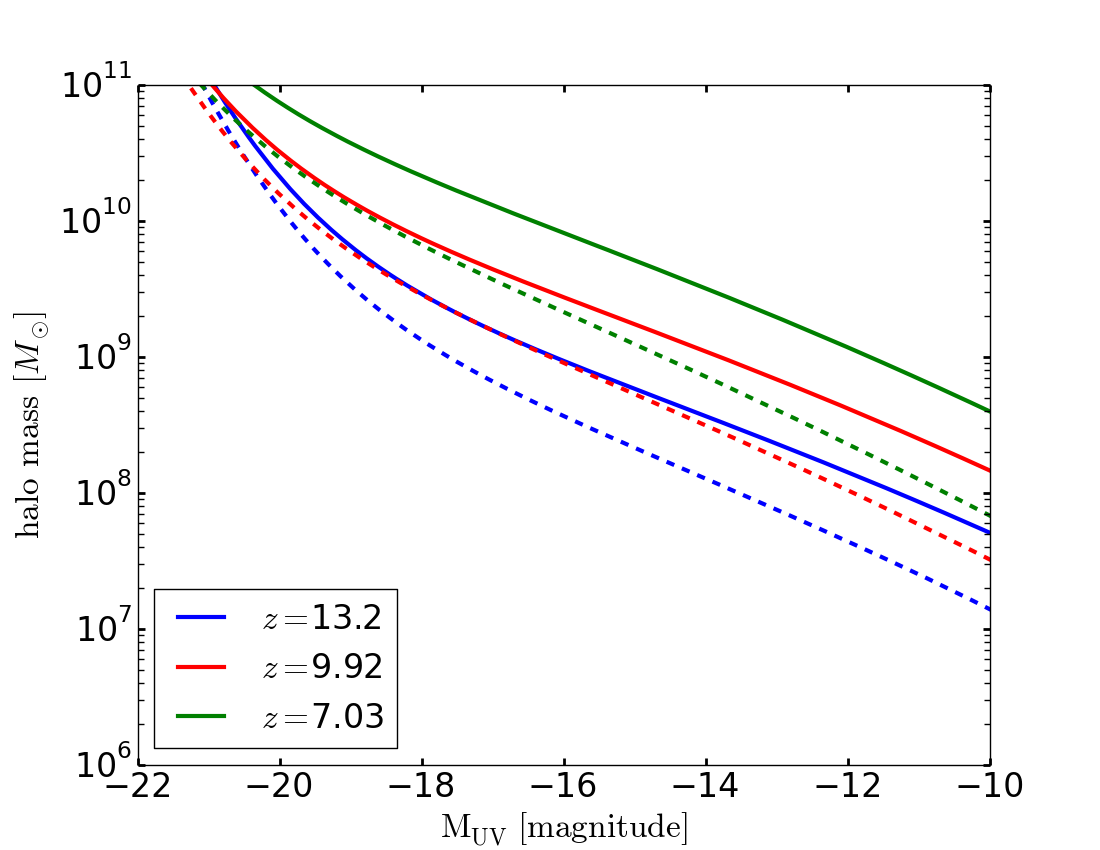}
\includegraphics[width=8.7cm]{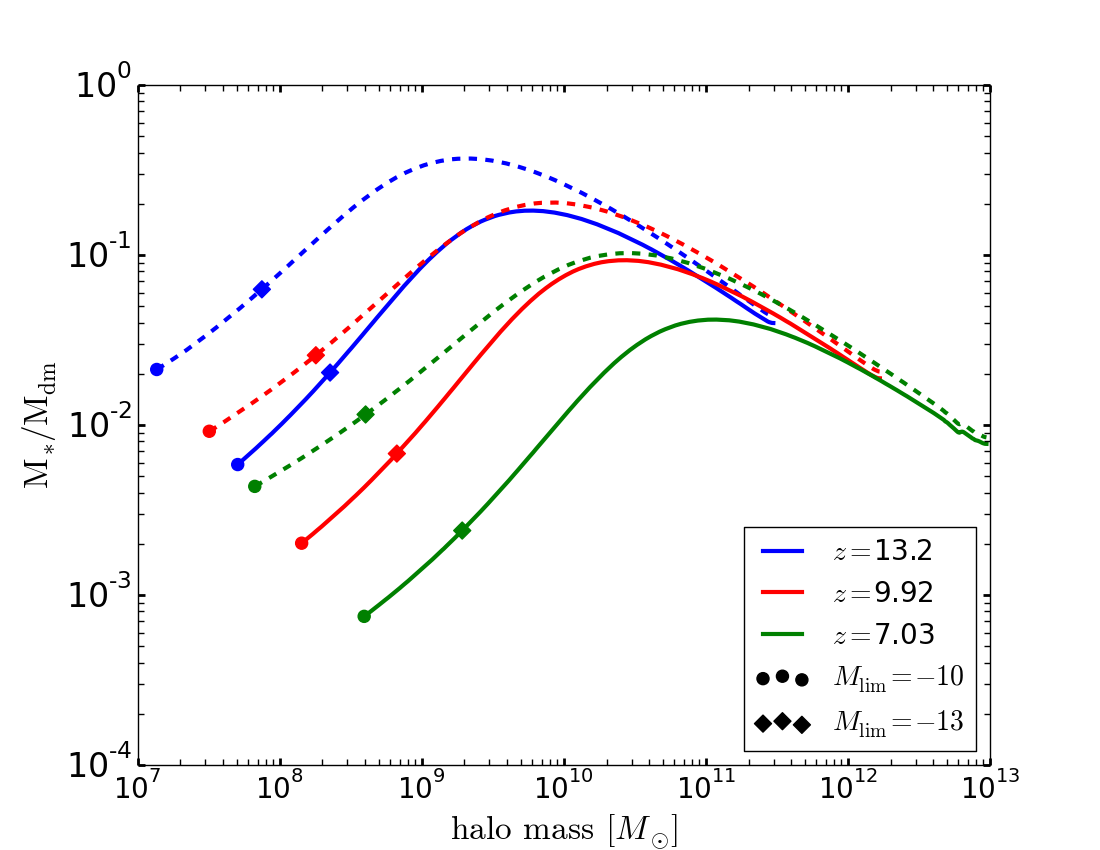}
\caption{({\it Left}). Typical halo mass hosting a galaxy of UV magnitude $M_{UV}$ produced by the halo matching method in \eq~\eqref{eq:halomatch}. The solid lines represent the continuous star formation, and the dashed lines represents bursty star formation with $f_{\text{duty}}=5\%$. ({\it Right}). Star formation efficiency $f_* \equiv M_*/M_{\rm dm}$ as a function of halo mass derived by our halo matching procedure. Halo matching in the bursty star formation models places brighter stellar populations in less massive halos.}
\label{fig:ML}
\end{figure*}
In Figure~\ref{fig:emiss} we plot the cumulative comoving hydrogen ionizing photon density derived from these models (blue, green, and red lines representing "FIT," "MAX," and "MIN," respectively) plotted alongside the total comoving baryon density of the universe (dotted lines) and $M_{UV,lim}=-13$. We have used the same definition for the conversion between UV magnitudes and ionizing photon luminosity $S_0$ as in \citet{Kuhlen:2012}:
\begin{equation}
S_0=2 \times 10^{25}~{\rm s}^{-1}\left(\frac{L_{\nu, 1500}}{{\rm ergs~s}^{-1} {\rm Hz}^{-1}}\right)\zeta,
\end{equation}
where $\log_{10}{(L_{\nu, 1500}/({\rm ergs~s}^{-1} {\rm Hz}^{-1}))}=0.4(51.63-M_{\rm UV})$.
The shaded regions represent the range of possible spectral hardness ($0.5<\zeta<2.0$). We interpret the crossing of the shaded regions with the dotted line as the earliest epoch at which the universe may be fully reionized (neglecting recombinations in the IGM). The addition of an escape fraction \fesc$<1$ reduces the total number of photons reaching the IGM. So, this figure can be used to infer the absolute lower limit for \fesc in the tree models we consider for the emissivity to reionize by $z_{re}=6$. Shifting the cumulative photon counts down by \fesc until the shaded region crosses the dashed line at $z=6$ gives \fesc for which each ionizing photon escaping into the IGM is used to ionize a hydrogen atom only once by redshift $z=6$.

Halos are matched so that in both continuous and bursty models the overall luminosity function is identical. However, in the case of bursty star formation, we assume a periodic UV luminosity with period $\Delta T$ with a simple step function functional form within each period:
\begin{align*}
L(t)=
\begin{cases}
L_{UV,1500}/f_{\rm duty}&0<t<f_{\rm duty}\Delta T\\
0&f_{\rm duty}\Delta T<t<\Delta T,
\end{cases}
\end{align*}
where $L_{UV, 1500}$ is the time averaged rest frame UV luminosity at $1500$~\AA\ over a period $\Delta T$, and the burst has peak luminosity $L_{UV,1500}/f_{duty}$ of duration $T_{\rm on}=f_{\rm duty}\Delta T$. We assign a luminosity $L(z)$ to a halo of mass $M(z)$ so that:
\begin{align}\label{eq:halomatch}
\int_{L(z)}^{\infty} \phi_L(L',z)dL'&=\int_{M(z)}^{\infty} f_{\text{duty}}\phi_M(M',z)dM'.
\end{align}
Here, we interpret $f_{\rm duty}\le 1$ as the fraction of halos emitting ionizing radiation at a given time. We thus take the number of available luminous halos for halo-matching to be $f_{\rm duty}\phi_M(M,z)$. In the case of continuous star formation, $f_{\text{duty}}=100\%$. 

In Figure~\ref{fig:ML}　(left) we show the result of this halo matching procedure at three sample redshifts indicated in the figure's legend. The UV luminosities are given in terms of UV magnitudes ($M_{UV}$) and the halo masses in solar masses. The solid lines represent the result of the procedure assuming continuous star formation ($f_{\rm duty}=100\%$), while the dashed lines represent the result of the procedure for bursty star formation with $f_{\rm duty}=5\%$. 
The figure shows that if galaxies are bursty, an observed galaxy at a given UV magnitude observed in HST deep fields lives in dark matter halo that is less massive than it would be inferred if their stars formed continuously.
The reverse argument is also true: a dark matter halo of a given mass that hosts a bursty galaxy has significantly higher UV luminosity (during the star burst) when compared to the luminosity of the same mass halo forming stars continuously. 
This also has indirect consequences on the theoretical expectations for \fesc in the continuous vs bursty models. Assuming that for a given mass halo the ISM structure is similar in the two models, the number of recombinations during the burst is proportional to the burst duration: $N_{\rm rec}=t_{\rm burst}/t_{\rm rec}$. The mean escape fraction over a burst cycle is \fesc=$1-N_{rec}/N_{\rm ph}$, where $N_{\rm ph}$ is the given total number of ionizing photons emitted in one cycle (which is same in the two models by construction). Thus, the escape fraction will be higher for shorter burst of star formation and the smallest for a continuous mode of star formation.
The right plot of Figure~\ref{fig:ML} shows the mean star forming efficiency $M_*/M$ (stellar mass per unit dark matter mass) as a function of halo mass for the same three sample redshifts. The symbols toward the small mass end of the curves show the typical halo mass of galaxies with UV magnitudes $M_{\rm UV, lim}=-10$ (circles) and $M_{\rm UV, lim}=-13$ (squares). This is going to be relevant for models in which the faint end of the luminosity function is extrapolated to $M_{\rm UV,lim}$ and assumed to be zero at fainter magnitudes. The figure shows that galaxies of a given total mass have higher star formation efficiency when assuming a bursty model instead of a continuous star formation model.

\subsection{Analytic approximation of Str{\"o}mgren spheres}

The starting point to derive our physical model for recombining \HII regions are the radiation transfer simulations presented in \citet{Ricotti:2001}. This paper presents 1D radiative transfer simulations around a point source of given spectrum (Population~II and Population~III stars or miniquasars) in an expanding universe following the ionization state of hydrogen, helium and the formation of H$_2$ via the H$^-$ catalyst.  We used the same code to generate the electron fraction around a halo with a given time dependent spectral energy distribution (SED). We derived simple analytic models of these outputs for both continuous and bursty star formation SEDs (see Appendix \ref{sec:approx}). The models allow to estimate the ionization fraction $x_e$ at a distance $R$ from a source or the distance from a source at which the electron fraction is $x_e$:
\begin{align}
x_e(R)&=f\left(z,R,L,z_{\rm on}(,z_{\rm off})\right)\\
R(x_e)&=g\left(z,x_e,L,z_{\rm on}(,z_{\rm off})\right),\label{eq:Rx_e}
\end{align}
where $L$ is the luminosity of the halo, $z$ is the redshift (independent variable), $z_{\rm on}$ is the redshift at which the star formation began, $z_{\rm off}$ is the redshift at which star formation ends in the case of bursty star formation. The analytic description of our physical model can be divided into two regimes:
\begin{enumerate}
\item During the burst of star formation the state of the IGM around an isolated halo is simulated as a cosmological Str{\"o}mgren sphere. The relatively low density of the IGM causes such spheres to have non-trivial transition regions. We found that the profile of these boundaries are well approximated by a simple analytic formula dependent only on a scale radius $R_S\left(z,L,z_0\right)$ (see Appendix \ref{sec:recomb}). The functional form of $R_S\left(z,L,z_0\right)$ is taken to be that of a Str{\"o}mgren sphere around a constant UV luminosity source in an expanding universe \citep[see,][]{Donahue:1987, ShapiroG:1987}.
\item After the burst of star formation: we assume that the ionization rate becomes zero and the gas recombines in the expanding universe. We solve analytically the equations for the evolution of the electron fraction under these assumptions at a given distance from the source as a function of time (see Appendix \ref{sec:ionize}). We take the state of the IGM at the moment when star formation ends as the initial condition for the electron fraction and gas temperature. Interestingly the results are quite sensitive to the assumed IGM temperature outside the \HII region, determined by the X-ray background radiation. With our analytic model we can therefore explore different scenarios beyond what we could do using pre-computed tables from a grid of 1D radiation transfer simulations.
\end{enumerate}
Our simulation models a cosmic volume by randomly placing halos within the volume and assigning to each a set of parameters $(L,z_{\rm on}(,z_{\rm off}))$ such that the luminosity functions at each redshift match observations at $z<10$ \citep[see,][]{Bouwens:2015} and extrapolations to $z>10$ \citep[see,][]{2008ApJ...688...85F, 2009ApJ...703.1416F, Kuhlen:2012}. The volume filling fractions at a given electron fraction $0<x_e<1$ (sampled uniformly with 60 bins in linear scale) are calculated using Equation~\eqref{eq:Rx_e} and output every 5 Myrs.

\subsection{Ionizing background}

The evolution of the IGM ionization fraction obtained using the method presented in the previous section approaches unity asymptotically as a function of cosmic time, but reionization is not fully complete (see Fig.~\ref{fig:Q}). Thus, the value of $\tau_e$ can be calculated quite precisely but the redshift of reionization remains undetermined. Methods which track the photon budget such as Equation~\eqref{eq:dqdt}, assume ionizing photons are absorbed instantaneously  and \HII regions have sharp boundaries. Our method allows for regions near ionizing sources to be fully ionized while large regions of gas further from the sources are only partially ionized. When the average electron fraction of the universe is small, our method (for continuous star formation) reproduces rather closely Equation~\eqref{eq:dqdt}, as photons are effectively absorbed locally and the mean free path is shorter than the typical distance between sources.
As the universe expands and the average electron fraction of the universe increases, the average density of neutral hydrogen decreases and the mean free path of photons in the IGM increases. These photons build up a ionizing background which becomes more dominant as more ionizing sources appear. The harder photons of the ionizing spectrum build up a background earlier than the softer photons due to their longer mean free path.
Individual halos never produce large enough regions where the gas is fully ionized to completely reionize the cosmic volume, so the derived average electron fraction underestimates the true electron fraction. We correct for this underestimation by calculating and including in the photon budget the ionizing background and its effect on the cosmic ionization history.

We quantify this effect solving the equation of radiation transfer in an homogeneous expanding universe as in \cite{Gnedin:2000, RicottiO:2004}, which we briefly summarize here. We begin with the number density of ionizing background photons $n_{\nu}$ at a redshift $z$ and evolve it to $z-\Delta z$. During each code timestep $\Delta z$, we add to the initial background at redshift $z$ (appropriately redshifted and absorbed by the neutral IGM) the photons produced by ionizing sources within our simulation between the redshifts of $z-\Delta z_0$ and $z-\Delta z$ including absorption and redshift effects (source term), where $\Delta z_0 \equiv H(z) R_0/c$ and $R_0$ is the minimum comoving distance of any emitting source that contributes to the radiation background. Mathematically, we solve the equation:
\begin{align}\label{eq:background}
&n_{\nu}(z-\Delta z)=n_{\nu}(z)\exp\left[-\int_z^{z-\Delta z}dz'\alpha_{\nu'}(z')\right]\nonumber\\
&\quad+\int_{z-\Delta z_0}^{z-\Delta z}dz' S_{\nu'}(z')\exp\left[-\int_{z'}^{z-\Delta z}dz''\alpha_{\nu''}(z'')\right],
\end{align}
where $\nu'=\nu (1+z')/(1+z)$ we have defined a dimensionless absorption coefficient and source function:
\begin{align}
\alpha_{\nu}&=\frac{(1+z)^2}{H(z)}c{\overline n}_{H}\sigma_{\nu}(\HI)(1-x_e(z)),\\
S_{\nu}&=\frac{{\dot n}_{\rm ion}\langle h\nu\rangle g_\nu/h\nu}{(1+z)H(z)},
\end {align}
where the sources spectra are normalized as $\int_{\nu_0}^\infty g_\nu d\nu=1$, with $h\nu_0=13.6$~eV and $\langle h\nu\rangle^{-1} \equiv \int_{\nu_0}^\infty (g_\nu/h\nu) d\nu$.
\begin{figure*}
\includegraphics[width=17.0cm,height=10cm]{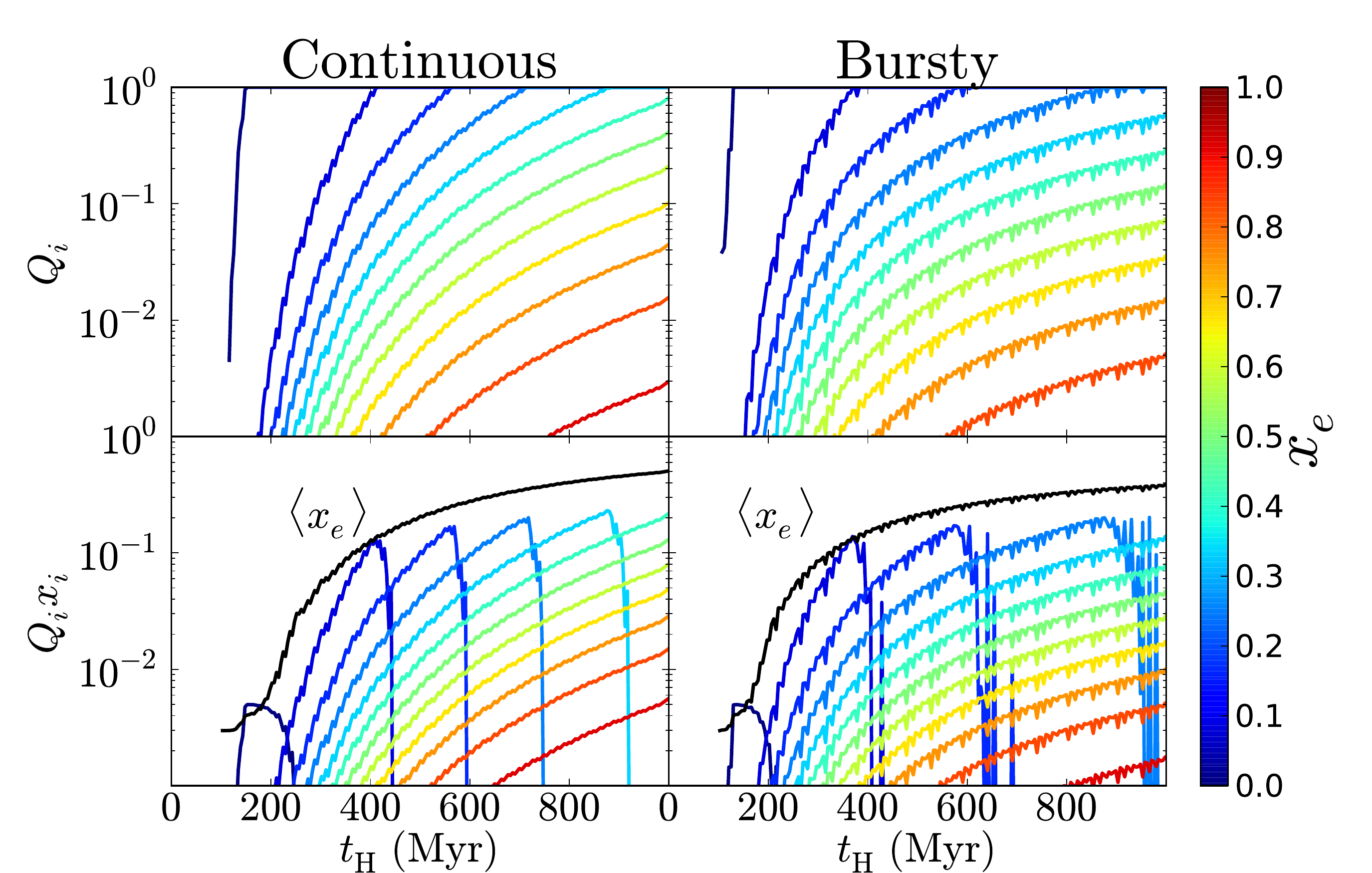}
\caption{({\it Top panels}). Volume filling fractions $Q_i$ of partially ionized gas with $x_e<x_i$ as a function of time for the continuous (left) and fiducial bursty (right) cases. ({\it Bottom panels}). Weighted electron filling fractions ($Q'_{i}x_i$) for the same models. The solid black lines represent the derived $x_e$ from \eq~\eqref{eq:meanchi} for both models. Note that a given electron fraction $Q'_{i}$ increase towards 1, but decrease as the next higher electron fraction $Q_{i+1}$ begins filling out the space indicated by $Q_{i}$.}

\label{fig:Q}
\end{figure*}
Sources at $R<R_0$ are local and their radiation is used to produce individual ionization bubbles and are thus excluded from the background calculation. We take $R_0$ to be the average distance between the dimmest (and most numerous) objects, so that the background begins at the distance of the nearest luminous objects. We also note that this distance decreases as more collapsed structures form, allowing the background to become more prominent at later times. The redshift of reionization $z_{\rm re}$ is rather sensitive to the choice of $R_0$ and the spectrum of the sources. Since it is difficult to make a precise estimate of $R_0$ (clustering of sources and other subtleties will affect the value of $R_0$) and because the frequency dependence of \fesc is unknown from either theory or observations, the redshift at which reionization is completed remains somewhat uncertain in our approximate separation between background and local sources. However, the relative difference in $z_{\rm re}$ between continuous and bursty models with different duty cycles is robust.

For our background calculation, we consider sources with a simple power law spectrum $g_\nu \propto \nu^{-1}$ truncated at $h\nu=100$~eV, absorbed by a column density $N_{\HI}$ of neutral gas. In order to consider the possibility that the preferential absorption of soft UV photons by neutral hydrogen inside the galaxy halo may result in a hardening of the spectrum emitted into the IGM, we consider two cases in our results: (i) $N_{\HI}=0$, {\it i.e.}, the spectrum is not affected by hydrogen absorption (ii) $N_{\HI}=3.8\times 10^{18}$~cm$^{-2}$, which corresponds to an escape fraction of \fesc$\sim 5\%$ and a significantly larger $\langle h\nu\rangle$ than in the case of an purely stellar spectrum. 
\hide{
\begin{figure*}
\includegraphics[width=17.0cm,height=10cm]{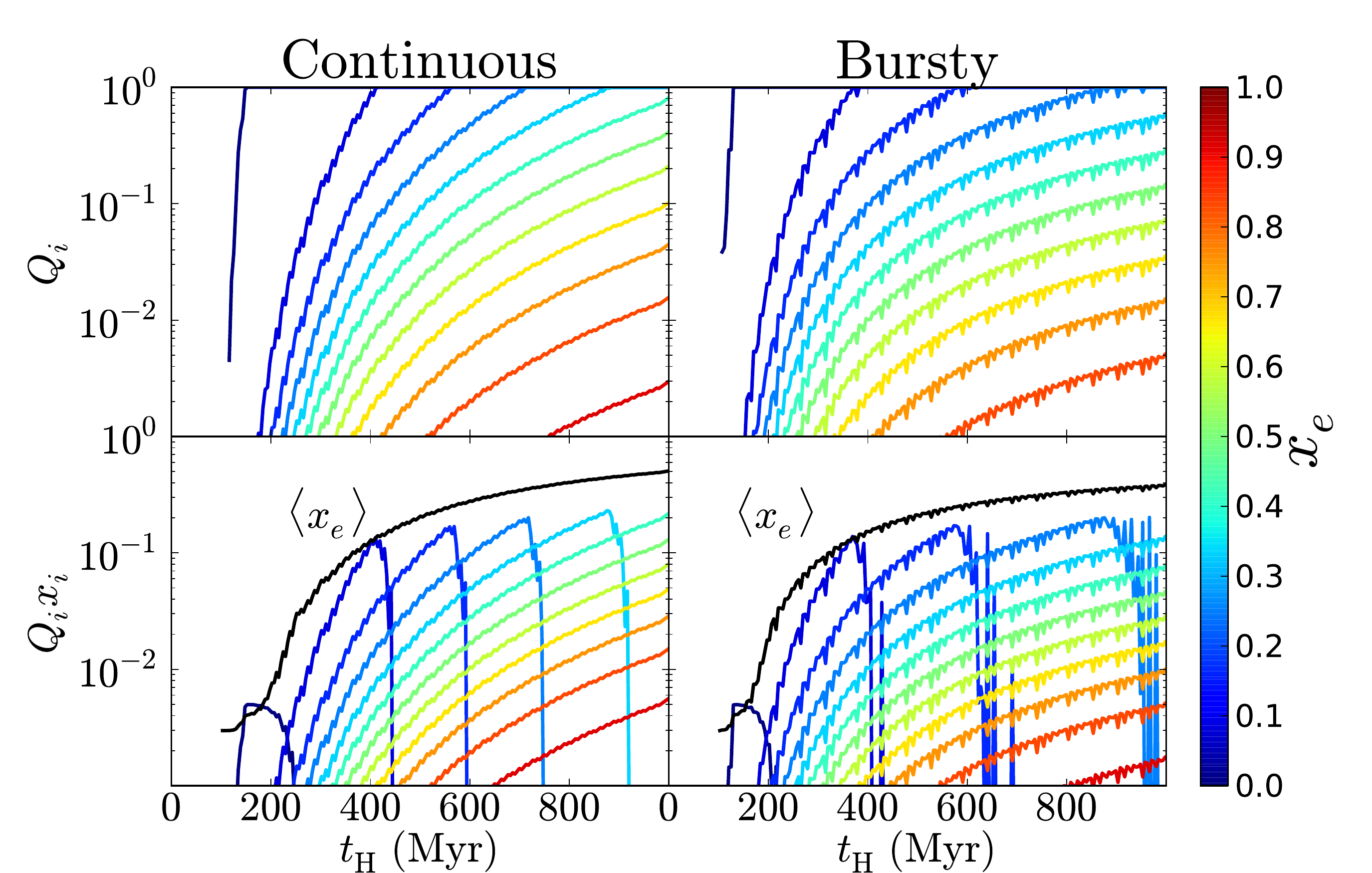}
\caption{({\it Top panels}). The raw output of the simulation for the continuous (left) and fiducial bursty (right) cases. ({\it Bottom panels}). Weighted electron filling fractions ($Q'_{i}x_i$) for the same models. The solid black lines represent the derived $x_e$ from \eq~\eqref{eq:meanchi} for both models. Note that a given electron fraction $Q'_{i}$ increase towards 1, but decrease as the next higher electron fraction $Q_{i+1}$ begins filling out the space indicated by $Q_{i}$.}
\label{fig:Q}
\end{figure*}
}
\subsection{Calculating observable quantities}\label{sec:taue}

Our simulation produces an array $Q_{i}(t)$ which represents the volume filling fraction of gas with electron fraction $x_e<x_i$ with $i=0,...,n-1$. We assume that regions of higher ionization are nested within regions of lower ionization, so that $Q_{0}<Q_{i}<...<Q_{(n-1)}$.
The filling fraction of gas with $x_{j-1}<x_e<x_j$ is
\begin{align}
Q'_{i}=
\begin{cases}
Q_{i}-Q_{i-1}&\text{if }1<i<n-1,\\
Q_{0}&\text{if }i=0.
\end{cases}
\end{align}
We compute the average electron fraction as:
\begin{align}\label{eq:meanchi}
\langle x_e(t) \rangle=\sum_{i=0}^{n-1}x_iQ'_{i}(t)
\end{align}
We let $\tau_e=\tau_0+\Delta\tau$, where $\tau_0$ is the contribution to $\tau_e$ from $z=0$ to the time our simulation ends ($z_0=5.8$) and reionization has happen, and $\Delta\tau$ is the contribution to $\tau_e$ from the simulation. The average optical depth of reionization may now be calculated using the formula:
\begin{align}
\Delta\tau_e(t)&=c\sigma_T\int_{t_0}^{t}dt\, n_e(t)=
c\sigma_T\int_{t_0}^{t}\left(\sum_{i=0}^{n-1}x_i n_H(t')Q'_{i}(t')\right)dt',\\
\tau_0&=\int_0^{z_0}dz\frac{c(z+z)^2}{H(z)}\sigma_T\left(1+\eta(z) T/4X\right).\label{eq:tau0}
\end{align}
Here, $H(z)$ is the Hubble parameter and $\sigma_T$ is the cross section of Thomson scattering. We let helium be singly ionized ($\eta=1$) for $z>4$ and doubly ionized ($\eta=2$) at lower redshifts so that we may directly compare our results with those presented in \cite{Kuhlen:2012}.

\section{Results}\label{sec:results}
\begin{figure*}
\includegraphics[width=8.7cm]{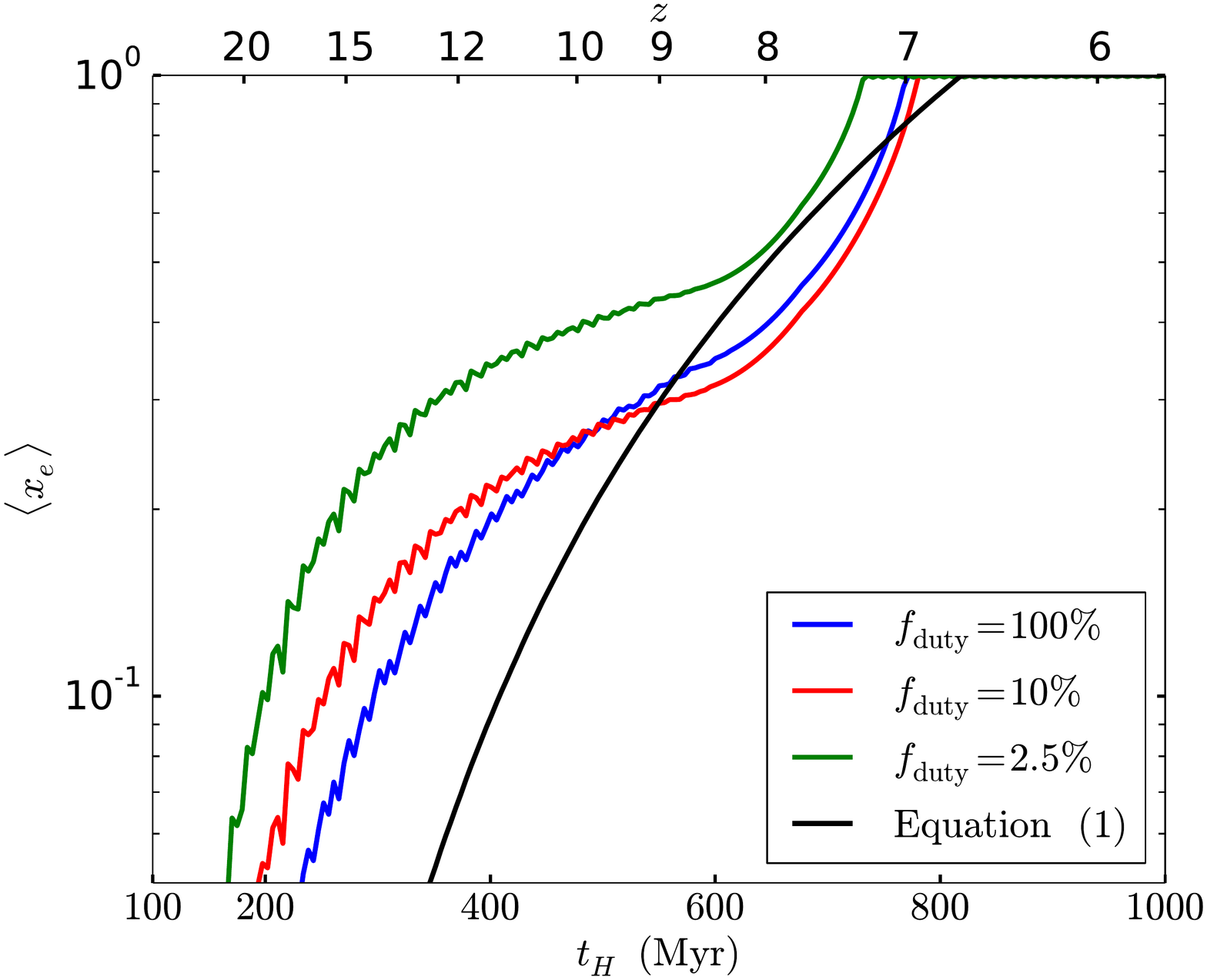}~\includegraphics[width=8.7cm]{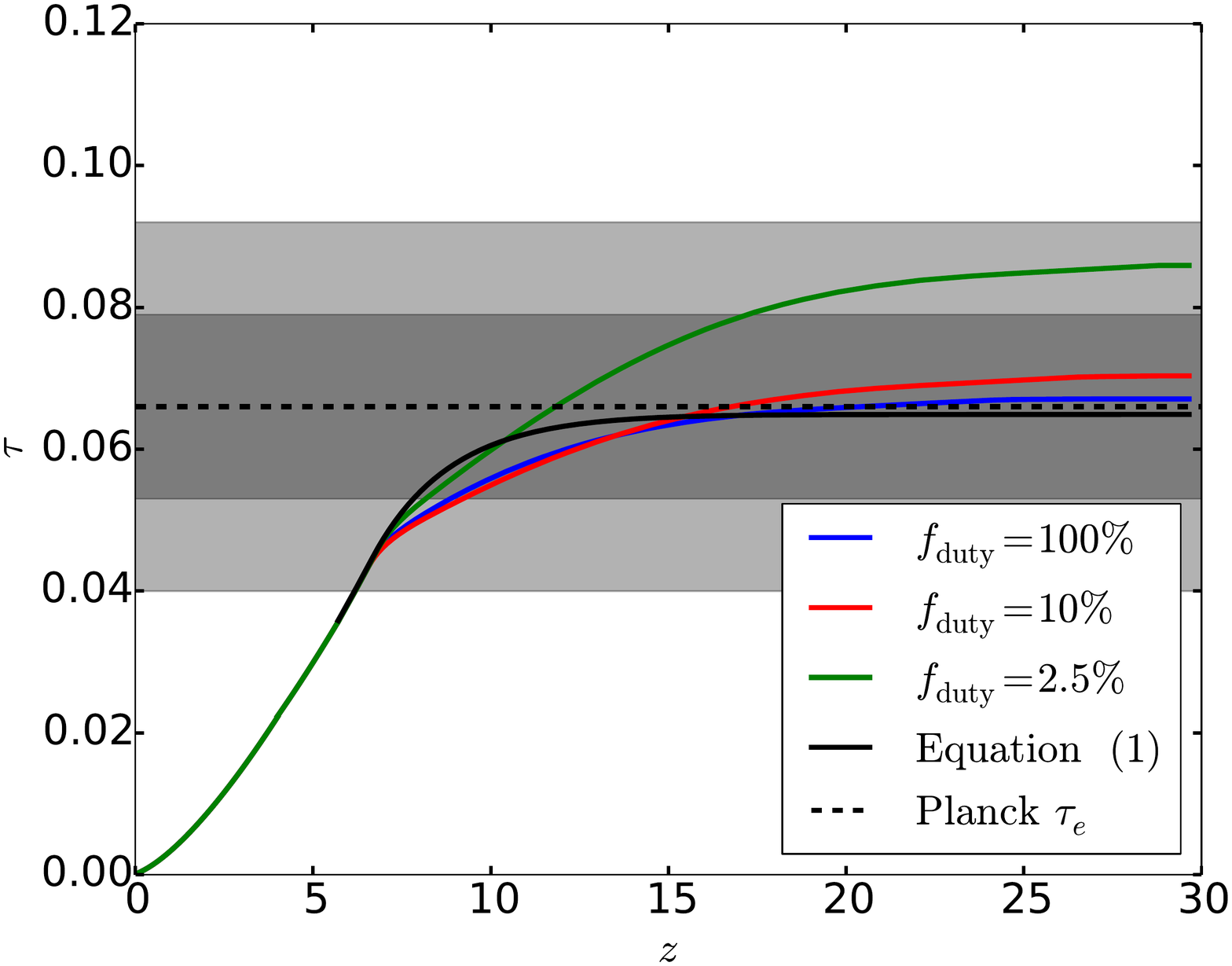}
\caption{Average electron fraction (left panel) and integrated $\tau_e$ (right panel) plotted as a function of time (and redshift) for the fiducial constinuous model and two bursty models ($f_{duty}=10\%$ red line, $f_{duty}=2.5\%$, blue line) with the same $\Delta T=100$ Mry and \fesc$=17\%$. The dashed line in the right panel shows the measured optical depth to Thompson scattering of the IGM due to reionization \protect{\citep{Planck:2015}}, with the shaded regions representing the 68\% and 95\% confidence regions. The black line in both plots represents a sample result produced by Equation~\eqref{eq:dqdt} with \fesc$=20\%$.}
\label{fig:xe}
\end{figure*}

For the results presented in the following sections, we chose UV luminosity functions with Schechter parameters from the FIT model. We take a fiducial choice of $M_{\text{UV, lim}}=-13$ and $m_{dm}=10^{7}$~M$_\odot$. For the instantaneous star formation case, we make fiducial choices of $\Delta T=100$~Myr and $t_{\rm on}=5$~Myr (thus, $f_{\rm duty}=5\%$). 
The top panels in Figure~\ref{fig:Q} show the primary output of our simulation, $Q_i(t) \equiv Q(x_e,t)$, for both continuous and instantaneous SF. We see that the filling fraction for higher levels of partial ionization are smaller in the instantaneous SF universe relative to those in the continuous SF universe. This is a result of the duty cycle of star formation in the case of instantaneous SF, wherein only the fraction of the halos actively producing stars are able to maintain high levels of ionization. We also see that the volume filling fraction at lower electron fractions rise at earlier times in the instantaneous SF case. This is due to the presence of relic regions of partial ionization around halos which have stopped forming stars and the higher luminosity of the burst at a given halo mass. The neutral-ionized transition regions surrounding star forming halos are not sharp for either star formation modes (see Appendix~\ref{sec:ionize} for a discussion), so there are large volumes of partially ionized \HII in both cases. These regions may be missed in full radiative transfer simulations which utilize only one or two frequency bins for ionizing radiation. 

The bottom panels of Figure~\ref{fig:Q} show the exclusive filling factors weighted by electron fraction, $Q'(x_e,t)x_e$, plotted for both continuous and instantaneous SF. We plot 20 of the 60 bins color coded such that higher electron fractions are more red and lower electron fractions are more yellow ($0.001<x_e<0.9$ in 20 uniform intervals as noted above). Each $Q'$ increases as more halos are formed until one of two things happen: (i) a large halo which dominates the region stops forming stars, in which case the higher electron fraction filling factors decrease more rapidly than the lower level filling factors; (ii) once a filling factor $Q(x_j,t)$ for a given electron fraction $x_j$ reaches unity: its corresponding exclusive filling factor $1-Q(x_{j+1},t)$ will only decrease as $Q$ for lower levels of ionization grow. The black lines in the bottom panels of Figure~\ref{fig:Q} show the average electron fraction $\langle x_e \rangle$, which is simply a sum of all of the curves below it.

The left plot of Figure~\ref{fig:xe} shows the average electron fraction as calculated with \eq~\eqref{eq:meanchi} for the continuous star formation and two choices of bursty star formation (both with $\Delta T=100$~Myr and $f_{duty}=10\%$ and $2.5\%$ respectively). We see that the electron fraction in the bursty case is higher at higher redshifts than in the continuous case. In the right panel of Figure~\ref{fig:xe} we plot the integrated optical depths to reionization as a function of redshift for the same models as in the left panel. For redshifts less than $z\sim 5.8$ at the end of our simulation, we integrated the formula for $\tau_e$ assuming a fully ionized universe as in Equation~\eqref{eq:tau0}. The shaded region shows the 1$\sigma$ and 2$\sigma$ confidence levels from \citep{Planck:2015}. Here we see that $\tau_e$ is significantly larger in the case of instantaneous SF even though $f_{\rm esc}$ and all other parameters are kept constant.

\subsection{Variation of parameters}
\begin{figure*}
\includegraphics[width=8.7cm]{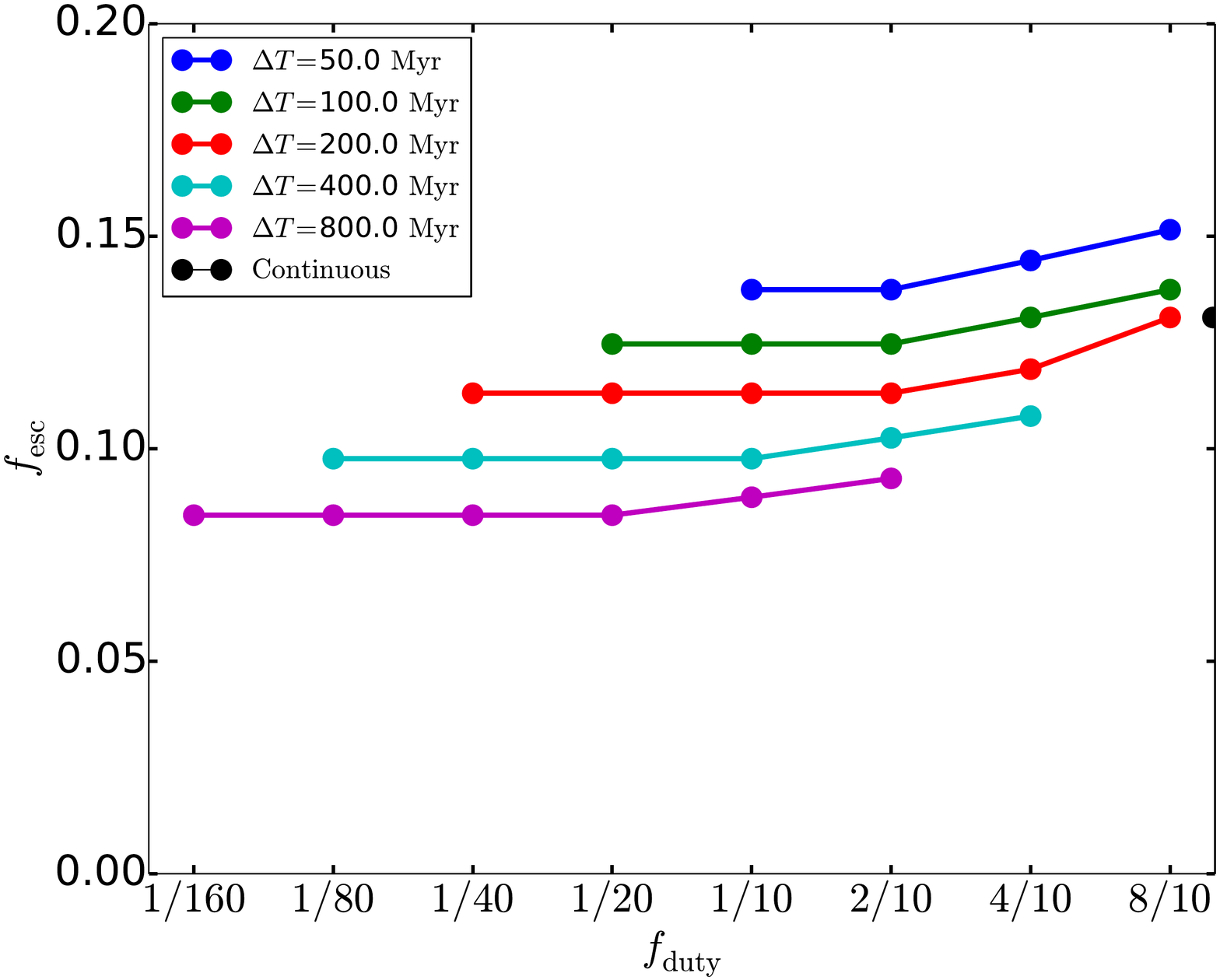}
\includegraphics[width=8.7cm]{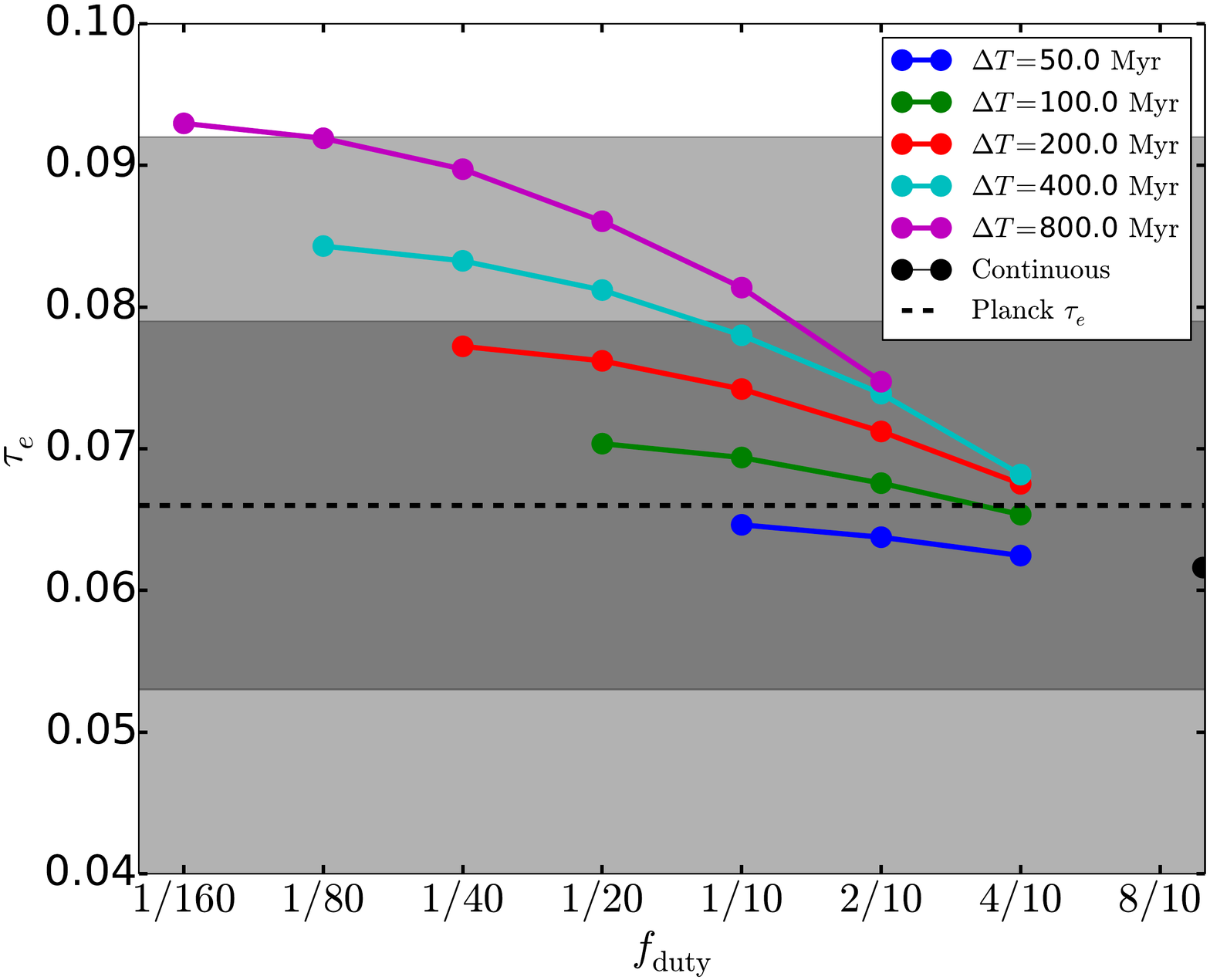}\\
\includegraphics[width=8.7cm]{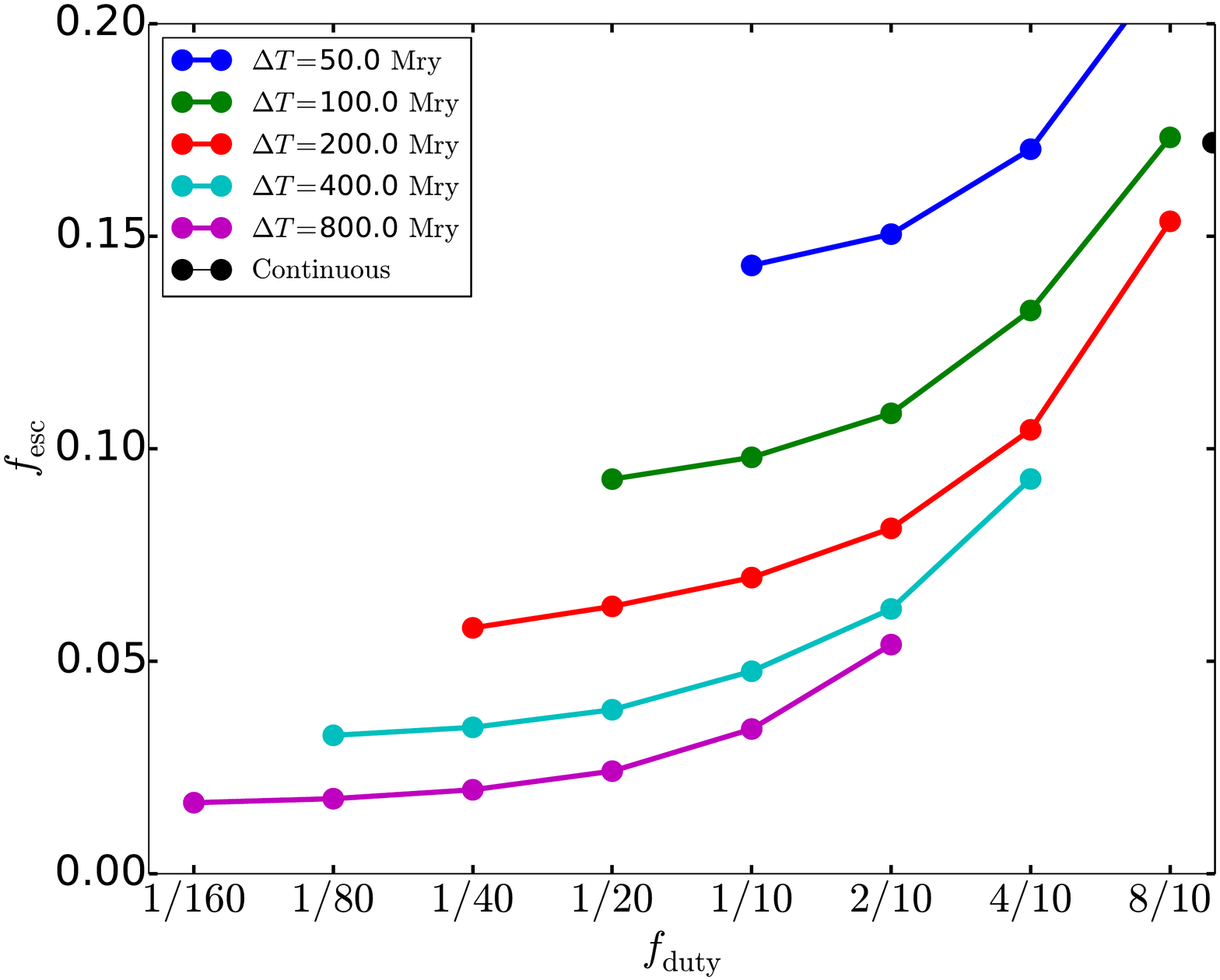}
\includegraphics[width=8.7cm]{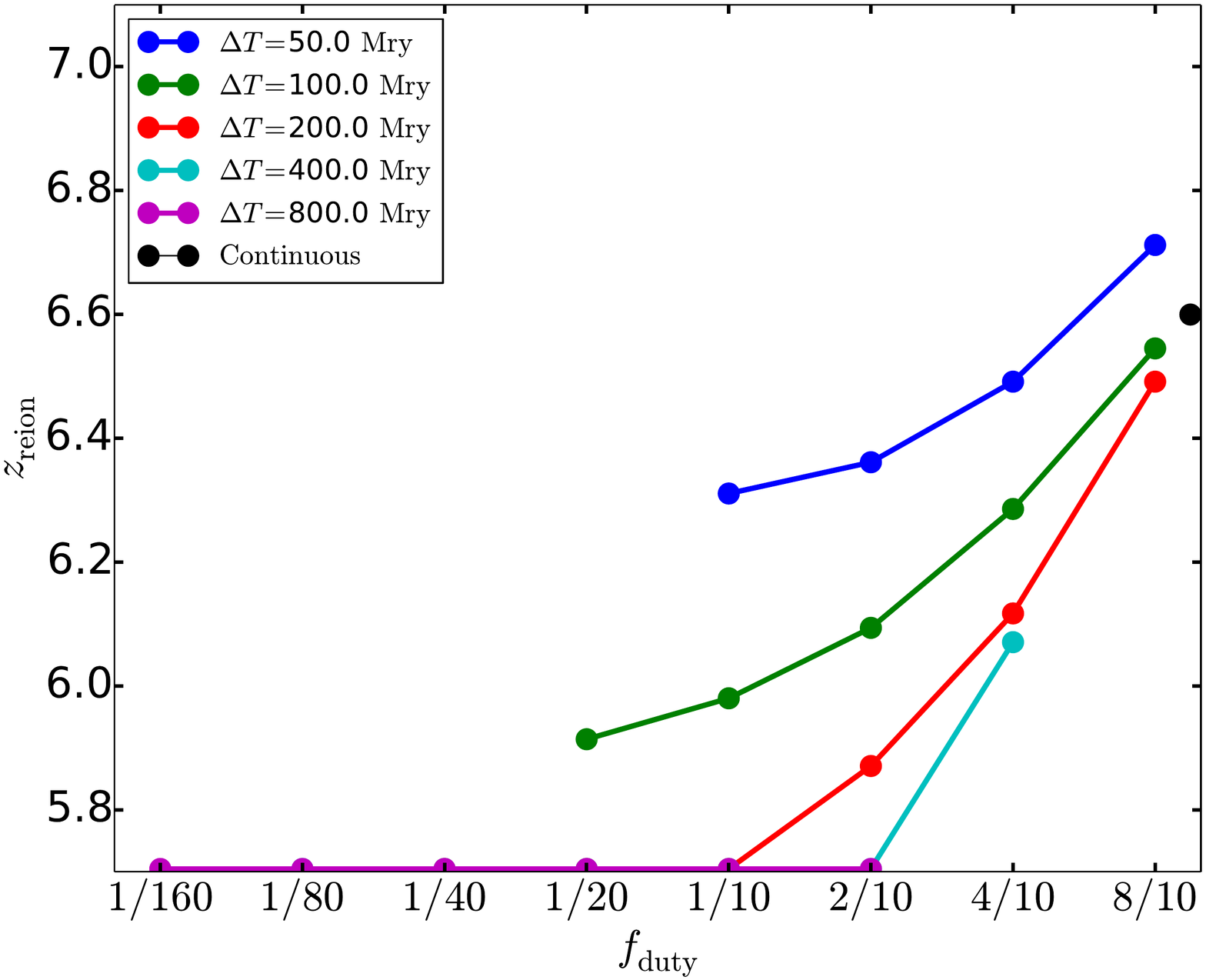}
\caption{Suite of runs for various values of $(\Delta T, T_{\rm on})$ matched to fixed observational constraints. All plotted points assume spectra produced within the halos are unaffected by absorption within the halos ($N_{HI}=0$~cm$^{-2}$). ({\it Top-Left}). Escape fraction $f_{esc}$ as a function of duty cycle, $f_{\rm duty}$, for five sets of runs completing reionization at $z_{\rm re}=6.25$.  Points connected by a line have the same burst periods $\Delta T$, as shown in the legend. We see here that runs with lower duty cycles require lower escape fractions to produce the same optical depth of reionization. ({\it Top-right}). IGM optical depth $\tau_e$ as a function of duty cycle for the same five sets of runs on the top-left. The dashed line and shaded regions refer to Planck's measurement of $\tau_e$ with 68\% and 95\% confidence intervals for the measurement error.
({\it Bottom-Left}). Escape fraction $f_{\rm esc}$ as a function of duty cycle, for the same five sets of runs but keeping the optical depth to Thompson scattering fixed at $\tau_e=0.066$. 
({\it Bottom-right}). Reionization redshift $z_{\rm re}$ plotted as a function of duty cycle for the same five sets of runs on the bottom-left. In all plots, the continuous case is represented by the black point at $f_{\rm duty}=1.0$.}
\label{fig:fduty1}
\end{figure*}

Our simulations allow us to compare the continuous star formation model to the bursty star formation model in a universe with freely chosen parameters including the burst duty cycle, $f_{\rm duty}$, burst period $\Delta T$, minimum halo luminosity $M_{\rm UV,lim}$, and minimum halo mass $m_{\rm dm}$. To help us understand the role of these parameters, we take either $\tau_e$ or $z_{\rm re}$ as observational constraints and vary $f_{\rm esc}$ for a given set of starting parameters to match the desired observational constraint.
The results of this parameter study are shown in Figure~\ref{fig:fduty1}.

\begin{figure*}
\includegraphics[width=8.7cm]{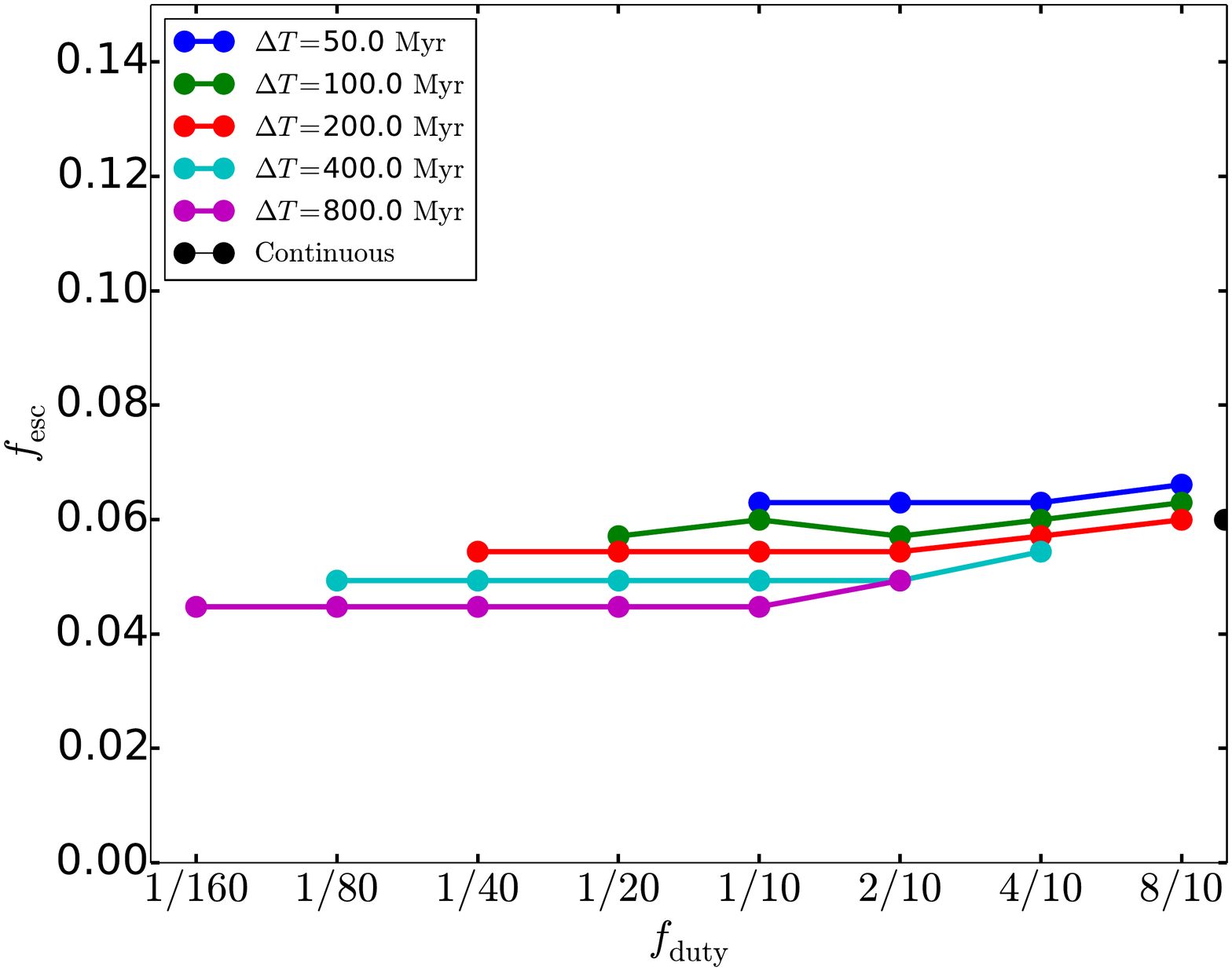}
\includegraphics[width=8.7cm]{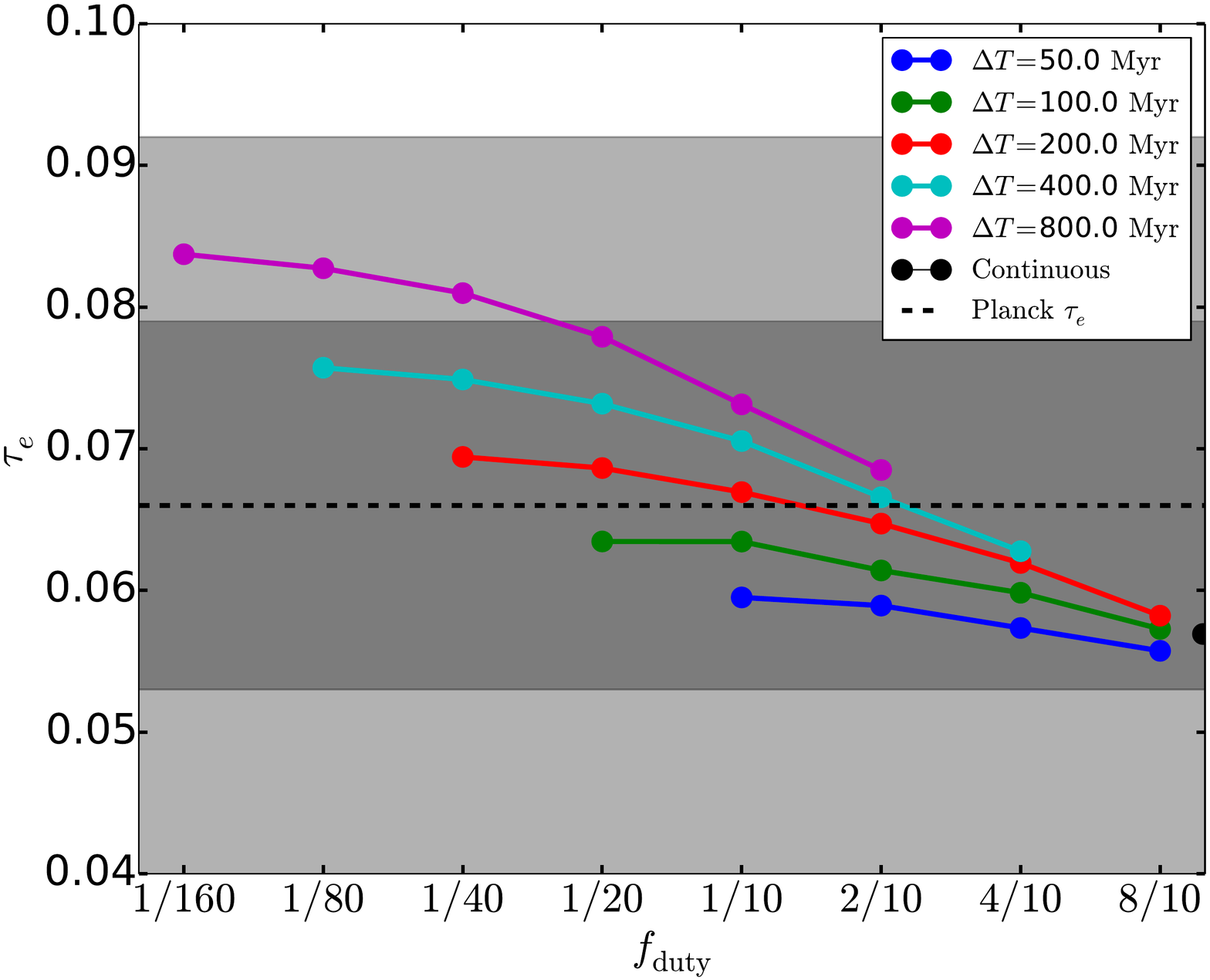}\\
\includegraphics[width=8.7cm]{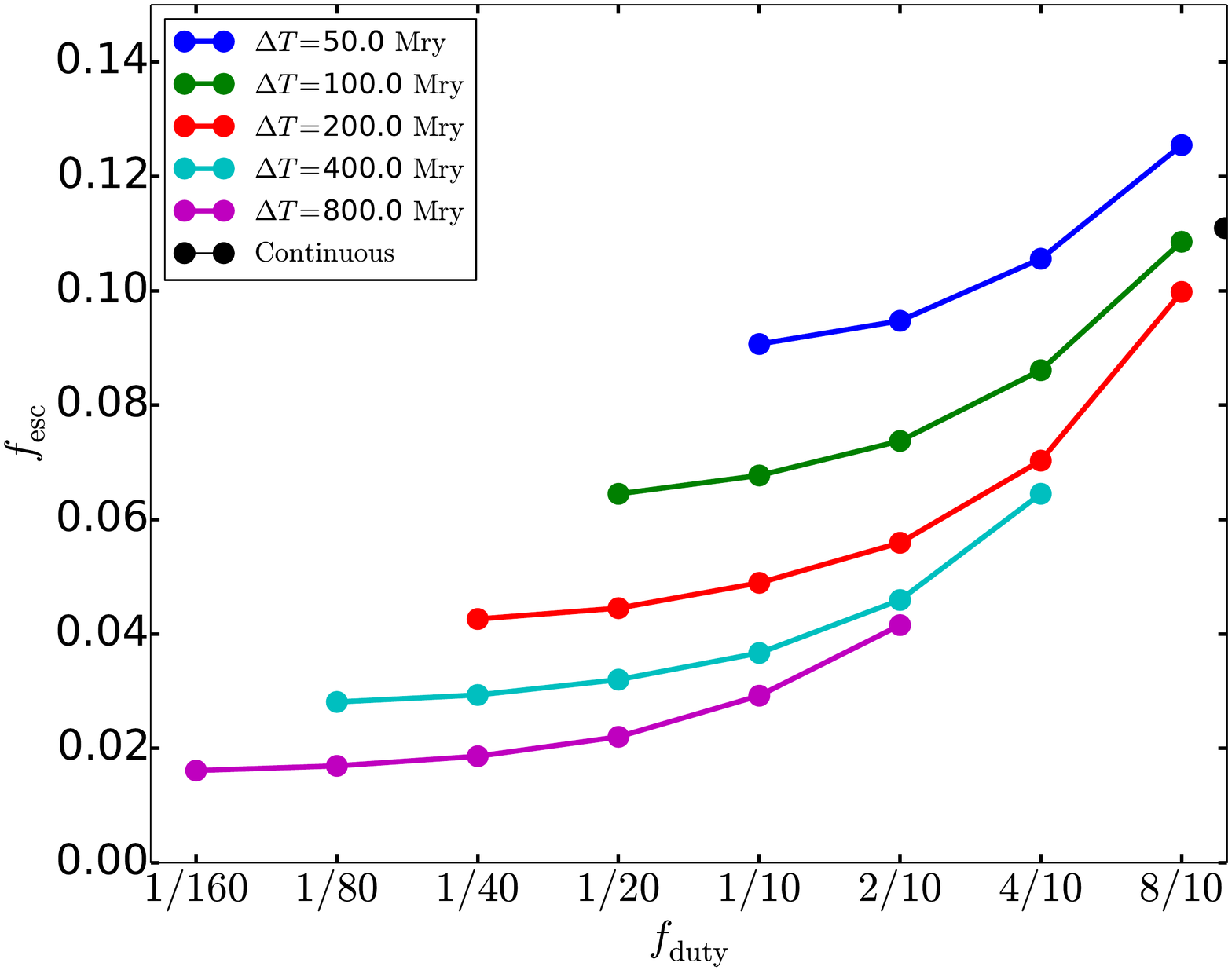}
\includegraphics[width=8.7cm]{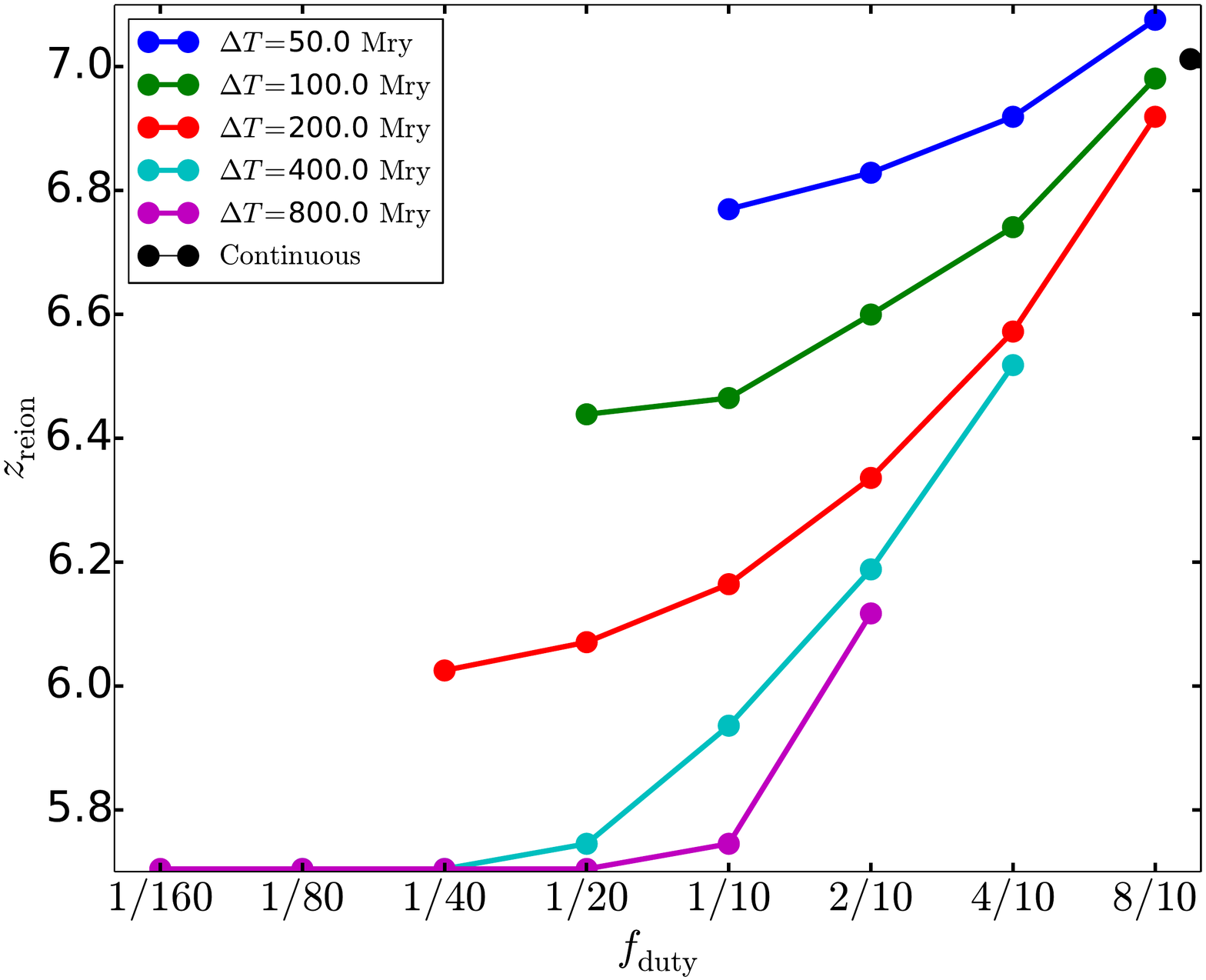}
\caption{Same plots as in Fig.~\ref{fig:fduty1} but assuming a spectrum of the sources absorbed by a fixed column density of neutral hydrogen $N_{HI}=3.8 \times 10^{19}$~cm$^{-2}$. Neutral hydrogen preferentially absorbs softer photons, thus resulting in a harder final spectrum. The overall result is a lower optical depth (for fixed $z_{\rm re}$), earlier completion of reionization (for fixed $\tau_e$), and a lower escape fraction (for fixed $\tau_e$, $z_{\rm re}$.)}
\label{fig:fduty2}
\end{figure*}

\begin{figure*}
\includegraphics[width=8.7cm]{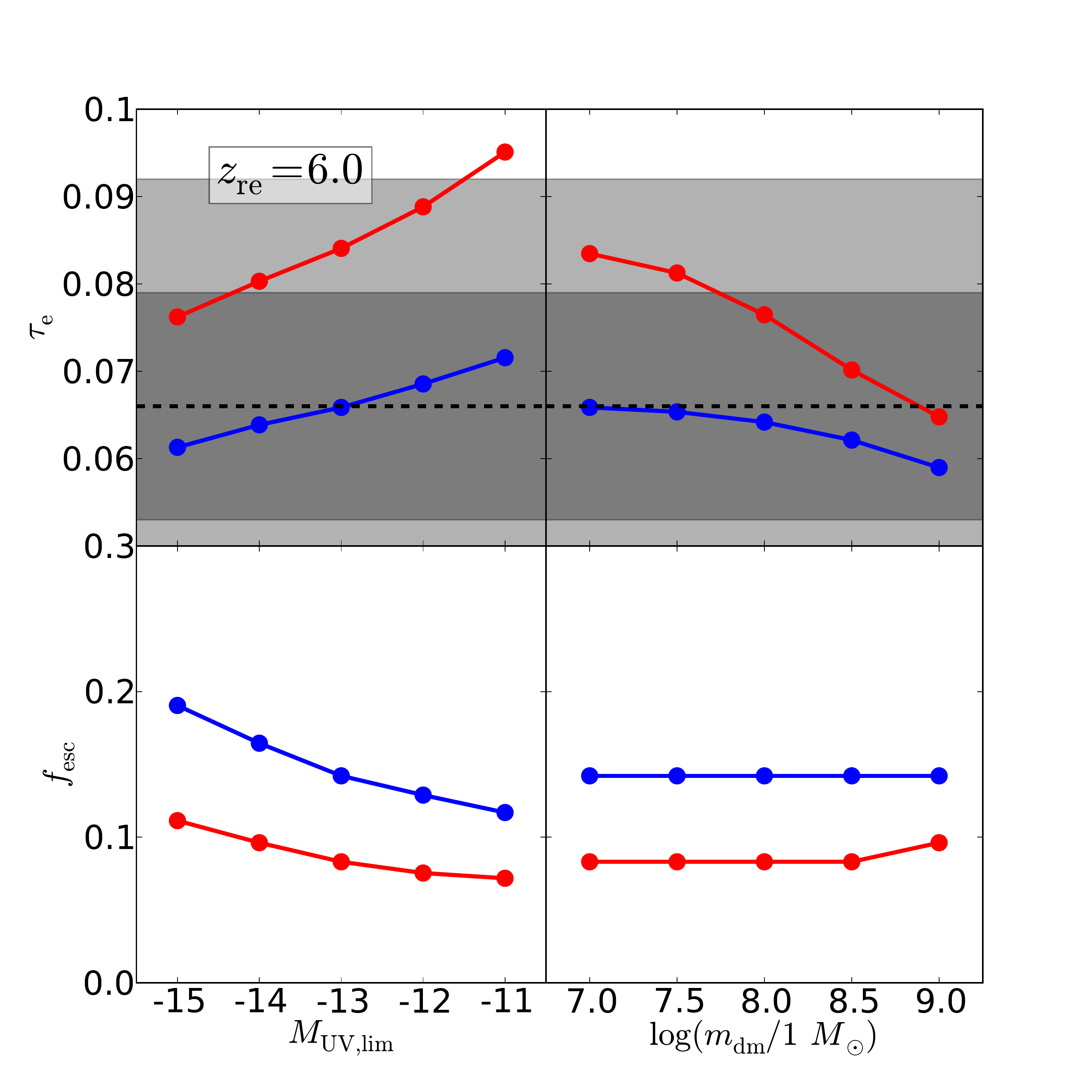}
\includegraphics[width=8.7cm]{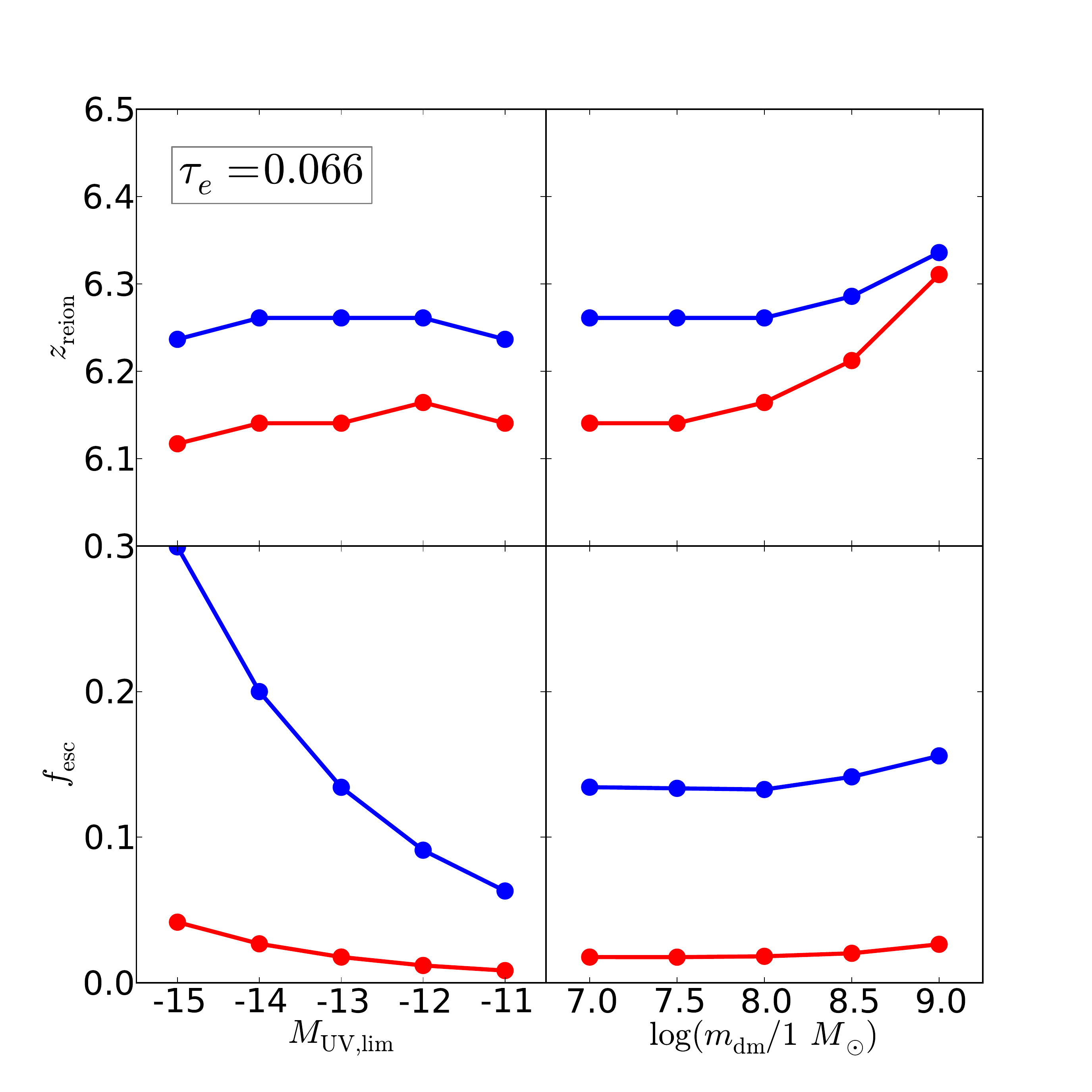}
\caption{Behavior of the complete model constrained to observable quantities under different choices of $M_{\rm lim}$ and $m_{\rm dm}$. The blue and red lines represent the continuous and bursty cases, respectively. For the bursty model we adhere to our original fiducial choice of $(\Delta T, T_{\rm on})=(50\,{\rm Myr},5\,{\rm Myr})$. ({\it Left}). Model constrained to $z_{\rm re}=6.0$. We see that increasing $M_{\rm lim}$, thus adding more photons to our simulation, increases $\tau_e$ while requiring a smaller $f_{\rm esc}$ to maintain $z_{\rm re}=6.0$. We also see that increasing $m_{\rm dm}$, thus removing lower mass halos from our simulation, decreases $\tau_e$ and increases $f_{\rm esc}$. Note that the bursty mode is more affected by this change, as in our simulation bursty stellar populations inhabit lower mass halos. ({\it Right}). Model constrained to $\tau_e=0.066$. We note that this plot exhibits similar trends as in the previous plot, with decreasing $z_{\rm re}$ replacing increasing $\tau_e$. We also note that the effect of altering $M_{\rm lim},\,m_{\rm dm}$ is greater in this case. This reflects the fact that our $z_{\rm re}$ is dominated by the ionizing background, and is thus less sensitive to variations in these parameters.}
\label{fig:mlimdm}
\end{figure*}

In the top panels we show a set of runs where we adjusted $f_{\rm esc}$ to match $z_{\rm re}=6.0$, allowing us to examine the effect of the burst period and duty cycle on the best fit $f_{\rm esc}$ (top-left panel) and resulting optical depth of reionization $\tau_e$ (top-right panel). We ran suite of simulations with $\Delta T$, $T_{\rm on}$ selected from the sets $\{50,100,200,400\}~{\rm Myr}$ and $\{5,10,20,40\}~{\rm Myr}$. $T_{\rm on}=5$~Myr was chosen as the shortest burst length as this is the timescale for the life of stars in a truly instantaneous starburst. Here we see that the duty cycle of the burst, that mainly affect the peak luminosity of the burst and the halo mass in which such luminosity is observed, does not have a strong effect on $f_{\rm esc}$ and therefore the redshift of reionization (that is constrained). This makes sense because the redshift of reionization is determined by the ionizing background that is sensitive to the average halo luminosity rather than its bursty nature. However increasing the period between bursts has the effect of reducing $f_{\rm esc}$ needed to reionize at $z_{rm re}=6$. This is likely because the enhance electron fraction of the IGM in bursty models makes the IGM transparent to ionizing radiation earlier. The increase of the electron fraction in bursty models with increasing $\Delta T$ is evident as an increase in $\tau_e$ shown in the top-right panel.

In the bottom panels we show a second set of runs where we constrained $f_{\rm esc}$ to produce $\tau_e=0.066$, this time to examine the effect of $\Delta T$ and $f_{\rm duty}$ on $f_{\rm esc}$ (bottom-left panel) and redshift of reionization $z_{\rm re}$ (bottom-right panel).

Here we see that decreasing the duty cycle decreases the required escape fraction to produce $\tau_e=0.066$. However, the decrease in $f_{\rm esc}$ is more sensitive to an increase of the period between bursts. Hence, the parameter that affects $\tau_e$ most significantly appears to be $T_{\text{off}} \equiv \Delta T - T_{\rm on}=\Delta T(1-f_{\rm duty}^{-1})$, allowing for the greater contribution from the relic regions of partial ionization to the ionization history. To summarize, we see that decreasing the duty cycle of bursty star formation (or increasing the period between bursts) produces the same $\tau_e$ with a lower required escape fraction and given that our results for continuous star formation agree with previous predictions found in the literature, we believe that this effect may alleviate the need for high escape fractions or extrapolations of the faint end of the luminosity function to very low values to explain the observed optical depth of reionization.

The results of Figure~\ref{fig:fduty1} assume a photon spectrum which is a pure stellar spectrum unaffected by absorption by neutral hydrogen in the ISM or gas within halos. In Figure~\ref{fig:fduty2}, we present the same results as Figure~\ref{fig:fduty1}, this time with a 
fixed absorption of a column density $N_{HI}=3.8 \times 10^{19}$~cm$^{-2}$ of neutral hydrogen. Lower energy photons have a shorter mean free path in neutral hydrogen, so that they are preferentially absorbed. The results presented in Figure~\ref{fig:fduty2} have the same total photon count as before, but with a higher $\langle h\nu\rangle$ of the harder spectrum. The top plots, for which $z_{\rm re}=6.0$, show a lower escape fraction and lower optical depth of reionization than Figure~\ref{fig:fduty1}. The harder photon spectrum produces a stronger ionizing background, thus lowering the required escape fraction and resulting optical depth of reionization. The bottom plots, for which $\tau_e=0.066$, show a lower escape fraction and higher redshift of reionization than Figure~\ref{fig:fduty1}. The ionizing background dominates the redshift of reionization, so that the earlied biuld up of the ionizing background due to higher average photon energy of the sources anticipates the redshift of reionization $z_{\rm re}$.

We also ran two suites of simulations to examine the effect of altering the parameters $M_{\rm UV,lim}$ and $m_{\rm dm}$. For the the first suite of runs we fixed $\Delta T=50$~Myr, $T_{\rm on}=5$~Myr, $m_{\rm dm}=10^{7}$~M$_\odot$ and let the luminosity cut at the faint end of the luminosity function vary: $M_{\rm UV,lim}=(-15,-14,-13,-12,-11)$. For the second suite of runs we fixed $\Delta T = 50$~Myr, $T_{\rm on}=5$~Myr, $M_{\rm UV, lim}=-13$ and let the cut of the halo mass function vary: $\log{(m_{\rm dm}/1~M_\odot)}=(7.0,7.5,8.0,8.5,9.0)$. In Figure~\ref{fig:mlimdm} we plot the results of these simulations, with $\tau_e$ and $z_{\rm re}$ constrained as in Figures~\ref{fig:fduty1}-\ref{fig:fduty2}. In both figures, the blue lines represent the continuous model, while the red line represents the bursty model.

The left plot of Figure~\ref{fig:mlimdm} shows the results of both suites with the constraint $z_{\rm re}=6.0$. We see that increasing $M_{\rm UV,lim}$, thus increasing number of halos emitting ionizing photons in the simulation, has the expected of effect of decreasing the escape fraction necessary to produce the fixed reionization redshift. Similarly, we see that increasing $m_{\rm dm}$, thus decreasing the number of low mass halos in the simulation, increases the escape fraction necessary to match the constraint reionization redshift. In the left panel we note that the increase in $f_{\rm esc}$ is marginal, while the decrease in $\tau_e$ is significant, so that low mass halos contribute significantly to $\tau_e$, but not as significantly to the ionizing background and $z_{\rm re}$. We also note that the change in both quantities is more significant for the bursty case, a result of the fact that bursty stellar populations in our simulation inhabit lower mass halos.

The right plot of Figure~\ref{fig:mlimdm} shows the results of both suites, now subject to the constraint $\tau_e=0.066$. The results of this plot reflect those of the previous plot; increasing $M_{\rm UV,lim}$ (adding photons) or decreasing $m_{\rm dm}$ (adding halos) decreases the escape fraction necessary to reach the constrained $\tau_e$. We note that the changes in $f_{\rm esc}$ and $z_{\rm re}$ are more significant than in the left panel, a reflection of the fact that $\tau_e$ is more sensitive to these parameters. We also note that increasing $M_{\rm UV,lim}$ too much for this bursty model means that $\tau_e$ can be nearly or completely achieved before the background completes reionization, so that these models can't complete reionization within the span of our simulation.

\begin{table*}
\centering
{\bf Table 1}\\
Observational constraints.
\begin{tabular}{lcccc}
\hline\hline
Constraint&&&&\\
\# & Redshift & Constraint & Technique & References \\\hline
1a.& 5.9 & $Q_{\HII}>0.89$ & Dark Gaps in Quasar Spectra & \cite{2015MNRAS.447..499M}\\
2a. & 6.24-6.42 & $Q_{\HII}<0.9\,(2\sigma)$ & Ly$\alpha$ Damping Wing of Quasars & \cite{2008suba.prop...50O}\\
3a.& 7.0 & $Q_{\HII}<0.5$ & Clustering of Ly$\alpha$ Emitting Galaxies & \cite{2014ApJ...794....5T}\\\hline
1b.& 6.0 & $\dot{n}^{\rm com}_{\rm ion}<2.6\,(<2.6)$ & Ly$\alpha$ forest & BH07, SC10\\
2b.& 5.0 & $\dot{n}^{\rm com}_{\rm ion}=4.3\pm2.6\,(\pm2.6)$ & Ly$\alpha$ forest & BH07, SC10\\
3b.& 4.2 & $\dot{n}^{\rm com}_{\rm ion}=3.5\pm0.8\,(^{+2.9}_{-2.2})$ & Ly$\alpha$ forest & FG08, P09\\
4b.& 4.0 & $\dot{n}^{\rm com}_{\rm ion}=3.2\pm0.4\,(^{+2.2}_{-1.9})$ & Ly$\alpha$ forest & FG08, P09\\
5b.& 3.8 & $\dot{n}^{\rm com}_{\rm ion}=2.8\pm0.3\,(^{+1.8}_{-1.6})$ & Ly$\alpha$ forest & FG08, P09\\
6b.& 3.6 & $\dot{n}^{\rm com}_{\rm ion}=2.6\pm0.3\,(^{+1.7}_{-1.5})$ & Ly$\alpha$ forest & FG08, P09\\
7b.& 3.4 & $\dot{n}^{\rm com}_{\rm ion}=2.8\pm0.7\,(^{+2.5}_{-1.8})$ & Ly$\alpha$ forest & FG08, SC10\\\hline
\end{tabular}
\caption{(a) Observational Constraints on the reionization history of the universe (b) Observational Constraints on the high redshift ionizing emissivity. Total uncertainties for $\dot{n}^{\rm com}_{\rm ion}$, which include systematic effects due to the spectral shape of the observational UV background and the thermal history of the IGM, are shown in parenthesis. $\dot{n}^{\rm com}_{\rm ion}$ is shown in units of $10^{50}s^{-1}\,{\rm cMpc}^{-3}$. Constraints are taken from
\citet{2007MNRAS.382..325B}(BH07), \citet{2010ApJ...721.1448S}(SC10), \citet{2008ApJ...688...85F}(FG08), and \citet{2009ApJ...705L.113P}(P09).}
\label{tab:data}
\end{table*}

\subsection{Consistency with Ly$\alpha$ forest Observations}

We have thus far focused on how the calculated $\tau_e$ or $z_{\rm re}$ may be constrained for a model with fixed parameters $(\Delta T, T_{\rm on}, M_{\rm lim}, m_{\rm dm})$ by varying $f_{\rm esc}$. For any given set of these parameters, our model produces an average electron fraction $\langle x_e(z)\rangle$ and ionizing emissivity $\dot{n}^{\rm com}_{\rm ion}(z)$ at all times throughout our simulation. The value of these quantities may be estimated from to redshifts as high is $z=6-7$ using observations of the Ly$\alpha$ forest. We included a collection of such data in Table~\ref{tab:data}. We may compare these observations to the outputs of our simulation (which ends at $z\sim5.8$) to see how well our outputs compare to real data, and if the inclusion of partially ionized precursor of \HII regions and relic \HII regions improves the agreement with observations.

\begin{figure}
\includegraphics[width=8.7cm]{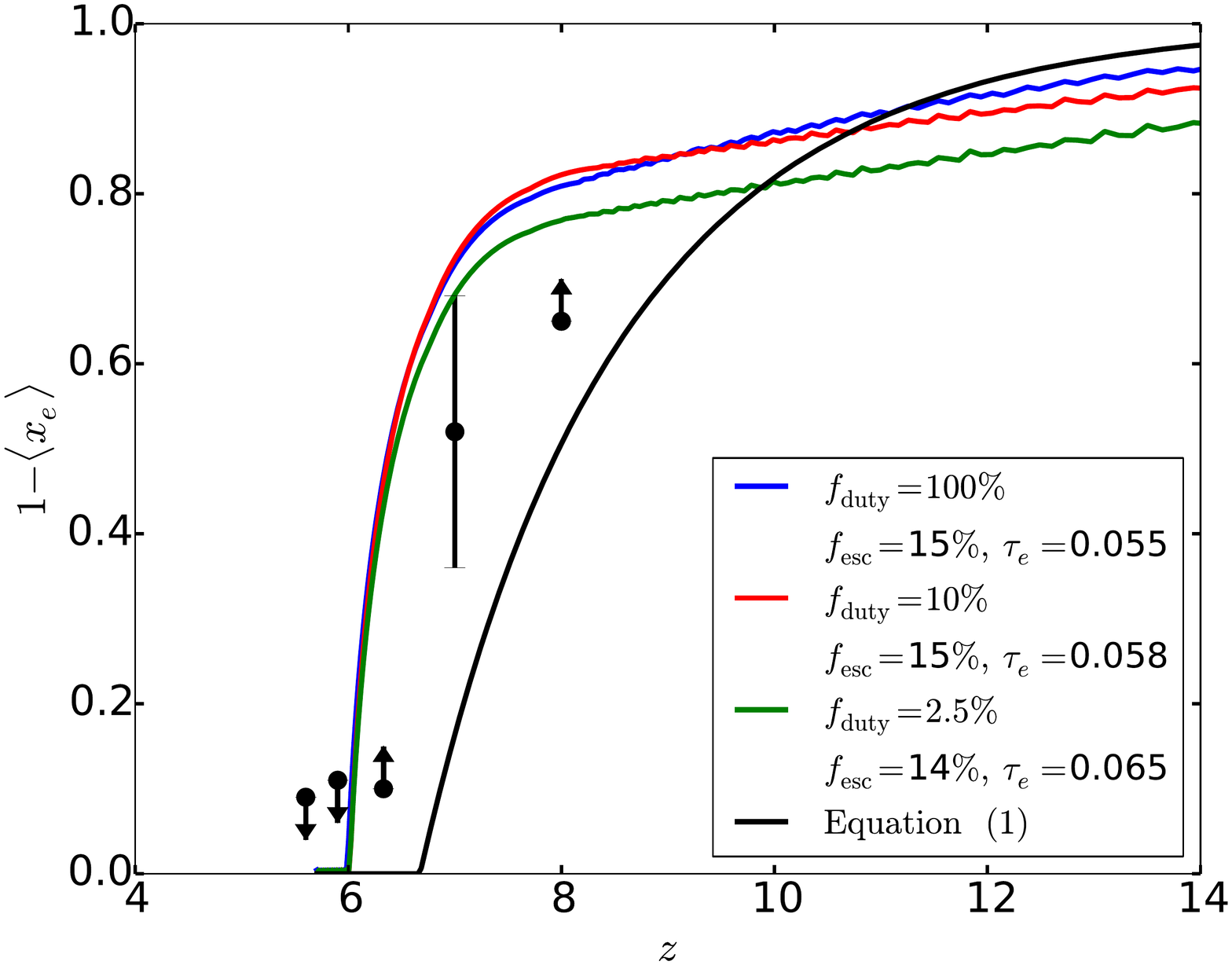}
\caption{A comparison of the simulated average electron filling fraction of the continuous model and two bursty models to observational data presented in Table~\ref{tab:data}. All three runs have $f_{\rm esc}$ adjusted so that $z_{\rm re}\sim6.0$. The black line represents a sample calculation using \eq~\eqref{eq:dqdt}. The ionizing background dominates when reionization is completed, and is most sensitive to the escape fraction, so that $f_{\rm esc}$ is similar for all three models. The bursty models, however, have higher electron fractions at high redshifts, so that $\tau_e$ is higher for these models.}
\label{fig:background}
\end{figure}

In Figure~\ref{fig:background} we plot the derived $\langle x_e\rangle$ for three simulations against the observational constraints on the average electron filling fraction $Q_{\HII}$ in Table~\ref{tab:data}. The model with $f_{\rm duty}=10\%$ has $(\Delta T, T_{\rm on})=(100,10)$~Myr while the model with $f_{\rm duty}=2.5\%$ has $(\Delta T, T_{\rm on})=(200,10)$~Myr. All three models have $f_{\rm esc}$ scaled as in the previous section so that $z_{\rm re}=6.25$ for best agreement with the data. The ionizing background dominates the completion of reionization, as can be seen from the sharp decrease in $1-\langle x_e\rangle$ around $z\sim 8$. The ionizing background is proportional to $f_{\rm esc}$, so that all three models have similar matched escape fractions (as in the fixed $z_{\rm re}$ plots of Figures \ref{fig:fduty1} and \ref{fig:fduty2}). The primary effect of decreasing the duty cycle is a boost in $\langle x_e\rangle$ at high redshifts, resulting in an increase in $\tau_e$, which we note in the figure.

\begin{figure}
\includegraphics[width=8.7cm]{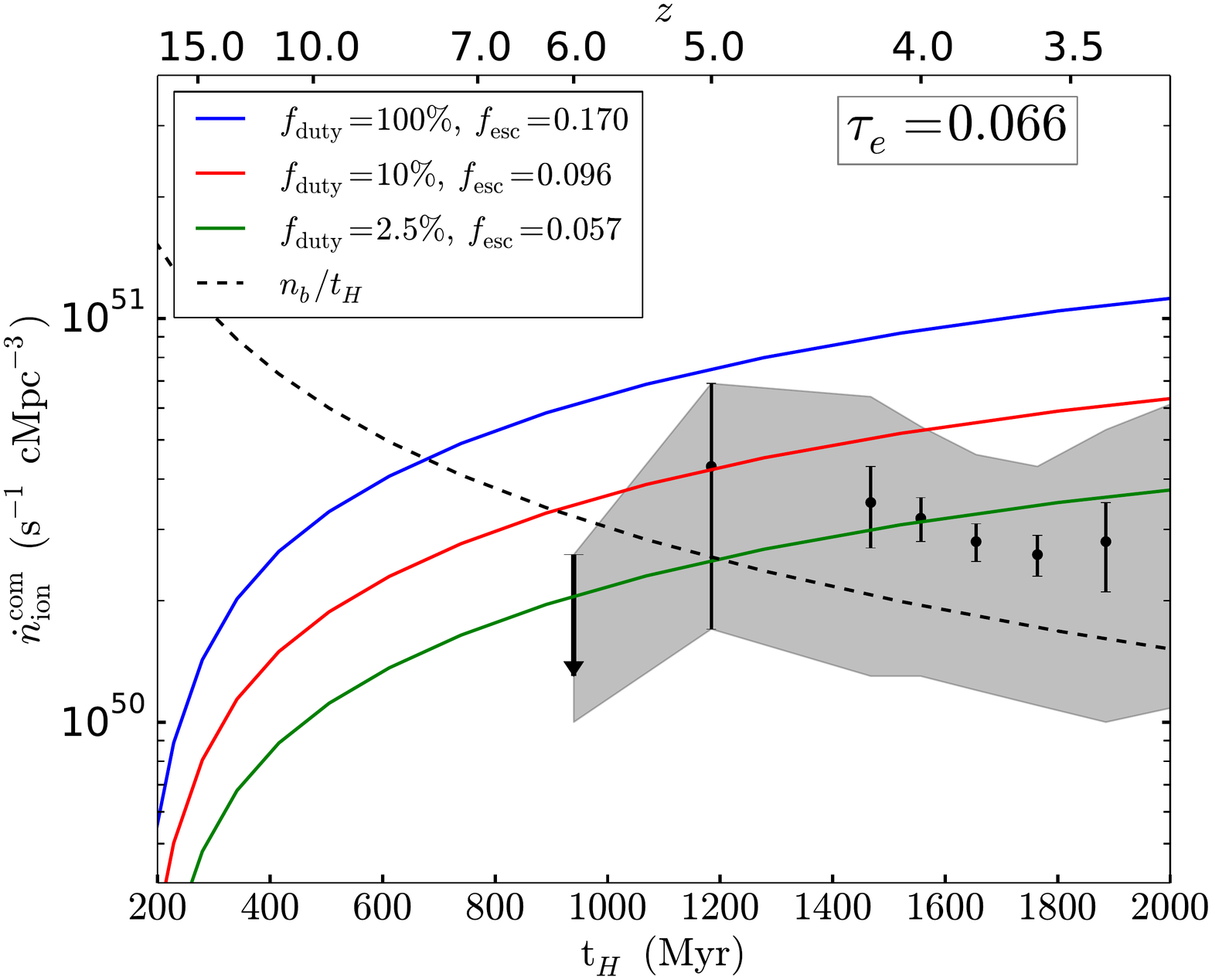}
\caption{A comparison of the spectrally integrated photon production rate for the continuous model and two bursty models with observational data presented in Table~\ref{tab:data}. All three runs have $f_{\rm esc}$ adjusted so that $\tau_e=0.066$. We see that a lower duty cycle allows for a lower escape fraction to produce the same $\tau_e$, allowing for better agreement with observational data.}
\label{fig:count}
\end{figure}

In Figure~\ref{fig:count} we plot the derived ionizing emissivity $\dot{n}^{\rm com}_{\rm ion}(z)$ for the same three models as a function of Hubble time against the observational constraints on this quantity shown in Table~\ref{tab:data}b. All three models have $f_{\rm esc}$ scaled as in the previous section so that $\tau_e=0.066$ to emphasize the effect of altering the duty cycle on $\tau_e$. The three models thus represent three points shown in Figure~\ref{fig:fduty1}; the escape fractions of the continuous, $f_{\rm duty}=10\%$, and $f_{\rm duty}=2.5\%$ models are $f_{\rm esc}=11.5\%$, $f_{\rm esc}=6.5\%$, and $f_{\rm esc}=3.9\%$, respectively. The short-dashed line shows $n_b(z)/t_H(z)$, the average number of baryons in the universe per Hubble time. The point at which the solid lines in each model crosses the short-dashed line gives a rough estimate of the lower bound for the redshift of reionization. The universe should be fully ionized once the average rate of ionizing photon production is a few times greater than $n_b/t_H$, which is in good agreement with our findings from the previous section. We see that continuous models must have an evolving \fesc, higher at high-z and decreasing toward lower redshifts in order to agree with the data points at $z<4$. An attractive feature of the bursty models is that a \fesc nearly constant with redshift agrees with observational data, because of the lower values of \fesc at $z>6$ needed to to produce $\tau_e=0.066$ and reionize by $z=6$. 

\section{Summary and Discussion}\label{sec:summary}
The model and analysis presented in this paper builds on the idea that bursty star formation in the early universe may increase significantly the average electron fraction of the universe when compared to continuous star formation, thus alleviating possible tensions between the observed high redshift UV luminosity functions and measurements of $\tau_e$. Stellar populations which form in bursts are luminous in the UV for about 5~Myr and mostly dark afterwards. Constraining models using high redshift UV luminosity functions and the halo matching method assuming bursty star formation, we found that the Str{\"o}mgren spheres which formed around halos of a fixed mass are larger than assuming continuous star formation, but begin to recombine once star formation ends. The recombination rate of ions in these halos is proportional to the square of the electron fraction, so that relatively long lived relic regions of ionization will be left behind in the IGM. Our goal was to build models to test whether this effect would have a non-negligible contribution to the average electron fraction of the universe, thus altering predictions of the volume filling fraction of \HII regions, $Q_{\HII}(z)$, the Thompson optical depth $\tau_e$, and the redshift of reionization $z_{\rm re}$, all of which are constrained to varying accuracy by observation.

In Section~\ref{sec:simulation} and in the Appendix, we presented the details of the model we used to compare ionization histories in cosmic volumes containing continuous and bursty star formation. The semi-analytical model tracks the evolution of stellar populations within statistically generated dark matter halos uniformly distributed throughout a volume and constrained to reproduce observed and extrapolated UV luminosity functions at high redshift. The electron fraction as a function of time around a given halo hosting bursty or continuous star formation is derived from writing down and solving the basic physics equations calibrated using one-dimensional radiative transfer simulations presented in \cite{Ricotti:2001}. The analytic equations we derived in the continuous and bursty cases are used to calculated volume filling fractions of partially ionized IGM gas for a uniformly sampled array of electron fractions. To calculate the cosmic ionization history we also include the contribution of the ionizing background. These quantities allowed us to derived the average electron fraction $\langle x_e\rangle$ throughout the volume as a function of time, from which we were able to calculate the relevant observable quantities.

We presented an in-depth analysis of the behavior of the our model and its predictions in Section~\ref{sec:results}. Our model allows us to freely choose the length of the starburst, $T_{\rm on}$, the period of repeated bursts, $\Delta T$, the escape fraction of ionizing photons, $f_{\rm esc}$, the hardness of the ionizing spectrum, the lower limit of halo luminosity function, $M_{\rm UV, lim}$, and the lower limit of dark matter halo mass, $m_{\rm dm}$. We took $f_{\rm esc}$ as an adjustable parameter to constrain either $\tau_e$ or $z_{\rm re}$ keeping fixed each time the other parameters, but exploring a range of possible parameter combinations.

The main results of this study are the following:
\begin{itemize}
\item By performing halo matching with the assumption of a duty cycle of star formation ($f_{\rm duty}=T_{\rm on}/\Delta T$), we find that the dark matter halos hosting galaxies of a given magnitude in the HST deep fields are less massive than they would be if we assume that stars formed continuously.
\item We speculate that the inferred smaller masses of these halos and the short duration of the luminosity burst should result in a larger fraction of the ionizing radiation produced by the stars escaping into the IGM.
\item For a fixed halo mass, stellar populations forming in the bursty mode are more luminous and produce larger \HII regions than in models where star formation is continuos. The relatively long-lived relic regions of partial ionization left behind by these bursts are able to maintain partial ionization throughout much of the IGM in a manner similar to X-ray pre-ionization, but heating the IGM less efficiently than X-rays. The overall effect is an increase in $\langle x_e\rangle$ at high redshifts, resulting in an increase in $\tau_e$ and a slightly higher redshift of reionization $z_{\rm re}$.
\item By constraining the galaxy UV luminosity density in our simulation, we find that to produce the $\tau_e =0.066$ observed by Planck and complete reionization by redshift $z_{\rm re}\sim 6.0$, models with bursty star formation require an \fesc $\sim 2\%-10\%$ that is $2-10$ times lower than in continuous star formation models (\fesc$\sim 17\%-20\%$). 
\item The ionizing photon budget needed to reproduce the observed $\tau_e$ depends strongly on the duty cycle of star formation but also the temperature of the partially ionized IGM that affects the life time of relic \HII regions. Thus even a relative low intensity of X-ray background radiation sufficient to increase the temperature of the IGM but not its ionization fraction may have an important indirect effect on the life span of relic \HII regions and therefore $\tau_e$. 
\item The hardness of the ionizing spectrum and radiation background instead affects $z_{\rm re}$. The hardness of the spectrum for stellar sources is determined mainly by the dependence of \fesc$(\nu)$ on the frequency. We find that increasing the column density $\langle N_{\HI}\rangle$ of neutral gas within the halo seen on average by the sources, produce a higher mean energy of escaping photons. A more energetic photon spectrum has a greater mean free path in the IGM, allowing the ionizing background to become dominant at earlier times and complete the reionization process earlier.
\item Our results suggest that any shortcoming of the ionizing photon budget suggested from extrapolated observations of the UV luminosity functions at high redshift and low values of \fesc$\sim 5\%$ typically measured in local starbursts \citep{}, would be alleviated if reionization was driven by short bursts of star formation, perhaps relating to the formation of Population~III stars and compact star clusters such as globular clusters, as suggested by previous studies\citep{}. A non-evolving \fesc$(z)=5\%$ in our bursty model is consistent not only with local observations of \fesc, $\tau_e$ from Planck, redshift of reionization $z_{\rm re}\sim 6$ from quasars, but also with the ionizing photon emissivity between redshifts $z \sim 2$ and $6$ inferred from observation of the Lyman-alpha forest.
\end{itemize}
The present study is only the first step to asses the effects of a bursty mode of star formation on the reionization history and on the the properties of halos hosting high-redshift galaxies.
Our model is fundamentally very simple and has several limitations but also some advantages and important improvements even with respect to full 3D radiative transfer simulations.

3D radiative transfer in cosmological simulations is basically monocromatic: because of computational limitations one can only afford to consider one frequency band for \HI ionizing radiation \citep[sometimes also \GI, \GII ionizing bands are considered,~\eg.][]{Gnedin:2000, Gnedin:2014}.
This may lead to an underestimate of the width of cosmological ionization fronts and therefore miss large volumes of partial ionization around \HII regions, in addition to the relic \HII regions left behind in a bursty mode of SF. Our analytic model is instead calibrated to reproduce 1D radiation transfer simulations in which the radiation field is sampled with more than 400 logarithmically spaced frequency bins. We are therefore able to capture the true width of cosmological \HII regions that is significant in the low density IGM (see Figure~\ref{fig:fit}(left) of the Appendix: the radius of the \HII region defined where $x_e=90\%$ is $0.2$~Mpc, while the radius where $x_e$ drops to $20\%$ is $1$~Mpc; that means that the region of partial ionization around \HII regions extends at least 5 times further than the Str\"omgren radius and the volume is 125 times larger). This may explain why our continuous SF model requires a slightly lower \fesc to produce $\tau_e=0.066$ than that found in other semi-analytic models that do not take into account regions of partial ionization.

The main limitations of our treatment are: i) we neglect clustering of sources and therefore the topology of reionization and size distribution of \HII regions \citep{Furlanetto:2004}. Realistically, the redshift of reionization will have a variance along different lines of sight. ii) Our treatment of the radiation background is an approximation. This is because without full radiative transfer calculation and clustering of the sources it is difficult to precisely account for the fractional contribution of local sources vs background sources (in practice because of this somewhat artificial separation, photons in the background can be slightly underestimated or overestimated). The background calculation is important in determining the redshift of reionization $z_{\rm re}$ but affects the value of $\tau_e$ only weakly. This is because the electron fraction produced by individual sources of reionization neglecting the background reaches values of $\langle x_e\rangle \sim 80\%-90\%$ while the background completes reionization bringing $\langle x_e\rangle$ to unity with a sharp rise similar to a phase transition \citep{Gnedin:2000}. Therefore in our model the absolute value of $z_{\rm re}$ may not be accurate. However, the trend of $z_{\rm re}$ as a function of the various free parameters in the model are robust. 

Theoretical estimates of \fesc are very uncertain \citep[\eg,][]{Ricotti:2000,Gnedin:2008a,Gnedin:2008b,Wise:2009,Yajima:2011,Yajima:2014} and observations are only possible in a limited number of cases, particularly local starburst galaxies and Lyman brake galaxies at $z \sim 3$ that in most cases set upper limits $f_{\rm esc}\simlt 4\%-8\%$ \citep[\eg,][]{Hurwitz:1997,2001ApJ...546..665S, 2003MNRAS.342.1215F, 2006MNRAS.371L...1I, 2006ApJ...651..688S, 2007ApJ...668...62S, 2009ApJ...692.1287I, Vanzella:2010, 2011ApJ...736...41B,Nestor:2013}. Our study found that values of \fesc$\sim 5\%$ are consistent with the observed values of $\tau_e$ and $z_{\rm re}$, even assuming truncation of the faint end of the high-z luminosity functions at $M_{UV,lim}=-15$ or truncation of the dark matter mass function at $10^9$~M$_\odot$and consistent with direct or indirect measurements at $z=0$ and $z \sim 1.3-6$. We thus believe that bursty star formation, that is likely to be prevalent in the early universe, will alleviate claims of missing sources of reionization suggested by several previous models. This different results follows from  taking particular care in modeling partially ionized gas in front of \HII regions and in relic \HII regions, thus reducing the number of ionizing photons used up by hydrogen recombinations (therefore increasing the effective $\overline t_{\rm rec}$ in Equation~\eqref{eq:dqdt}). We therefore caution of simply using Equation~\eqref{eq:dqdt} to constrain the ionizing emissivity and \fesc from the observed $\tau_e$. The redshift of reionization $z_{\rm re}$ also cannot be used to constrain the local ionizing emissivity because it is determined by the radiation background emitted by sources not yet observable and is quite sensitive to the hardness of the sources spectra. 

\subsection*{ACKNOWLEDGMENTS}
MR thank the National Science Foundation for support under the
Theoretical and Computational Astrophysics Network (TCAN) grant
AST1333514 and CDI grant CMMI1125285. Thanks to the anonymous
referee.

\bibliographystyle{./mn2e}
\bibliography{paper01}

\appendix

\section{Analytic Approximations}\label{sec:approx}

\subsection{Ionization Profile During Star Formation}\label{sec:ionize}

A simple model for the shape of the ionization profile may be found by solving the following formula iteratively:
\begin{align}
x_e^2(u)=-\frac{1}{3u^2}\frac{d}{du}\exp\left(-\tau_0\int_0^u(1-x_e(u'))du'\right),
\end{align}
where $u=r/R_s$ is the dimensionless distance from the source and $R_s$ is a scale length of the Str\"omgren sphere and $\tau_0=n_{\rm HI}\sigma_\nu(\HI) R_s$ is the optical depth of neutral hydrogen for a column density of $N_{HI}=n_{\rm HI}R_s$. From this formula we see that $\tau_0$ represents a scale length in the $u$ domain for the drop in electron fraction. While this model is fairly precise, we require a simple analytic formula to construct our statistical model. We may, however, use the above formula to note that both since $\tau_0 \propto N_{HI}$, the width of the ionization profile in comoving coordinates should scale with $N_{HI}$.

\begin{figure*}
\includegraphics[width=8.7cm]{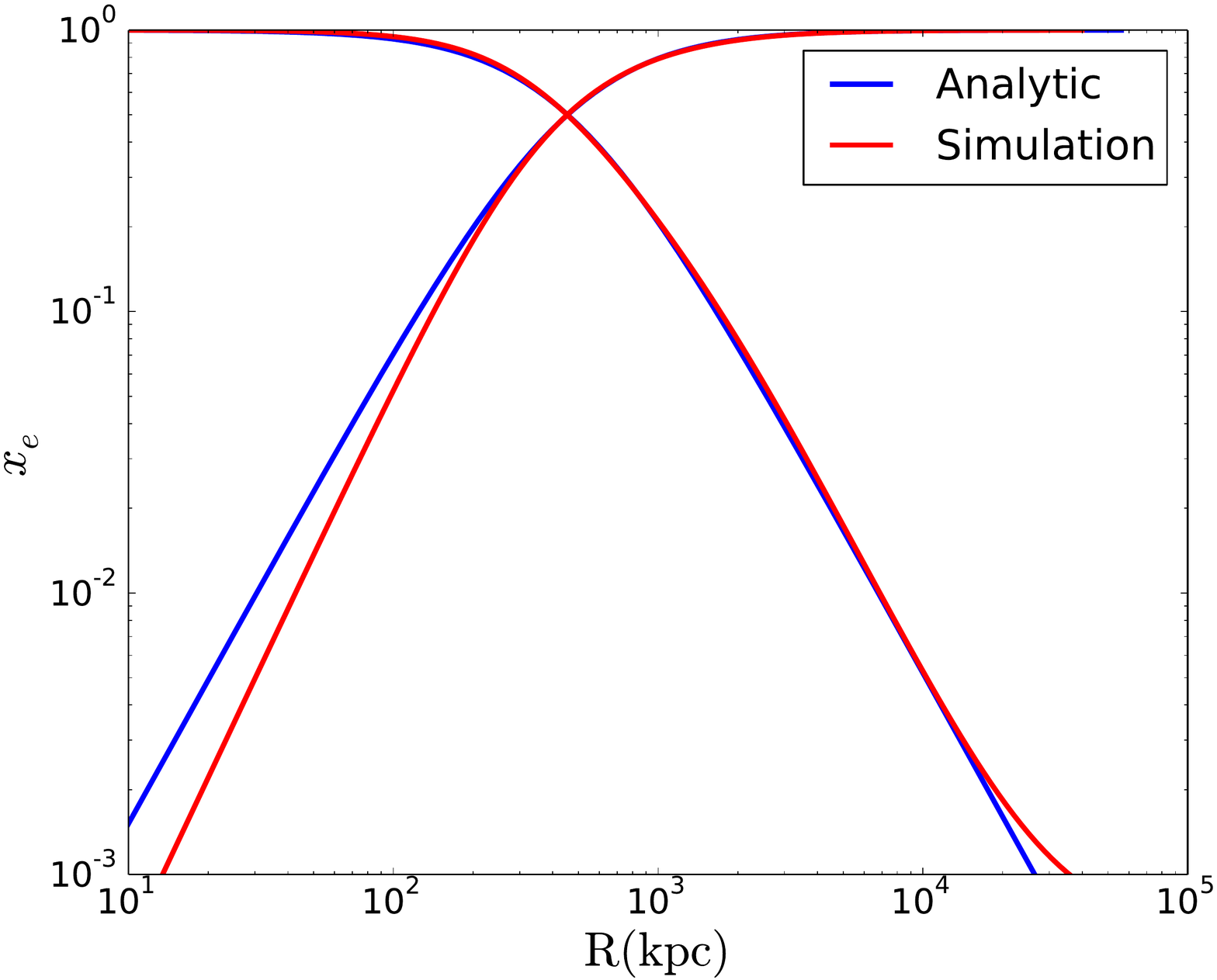}
\includegraphics[width=8.7cm]{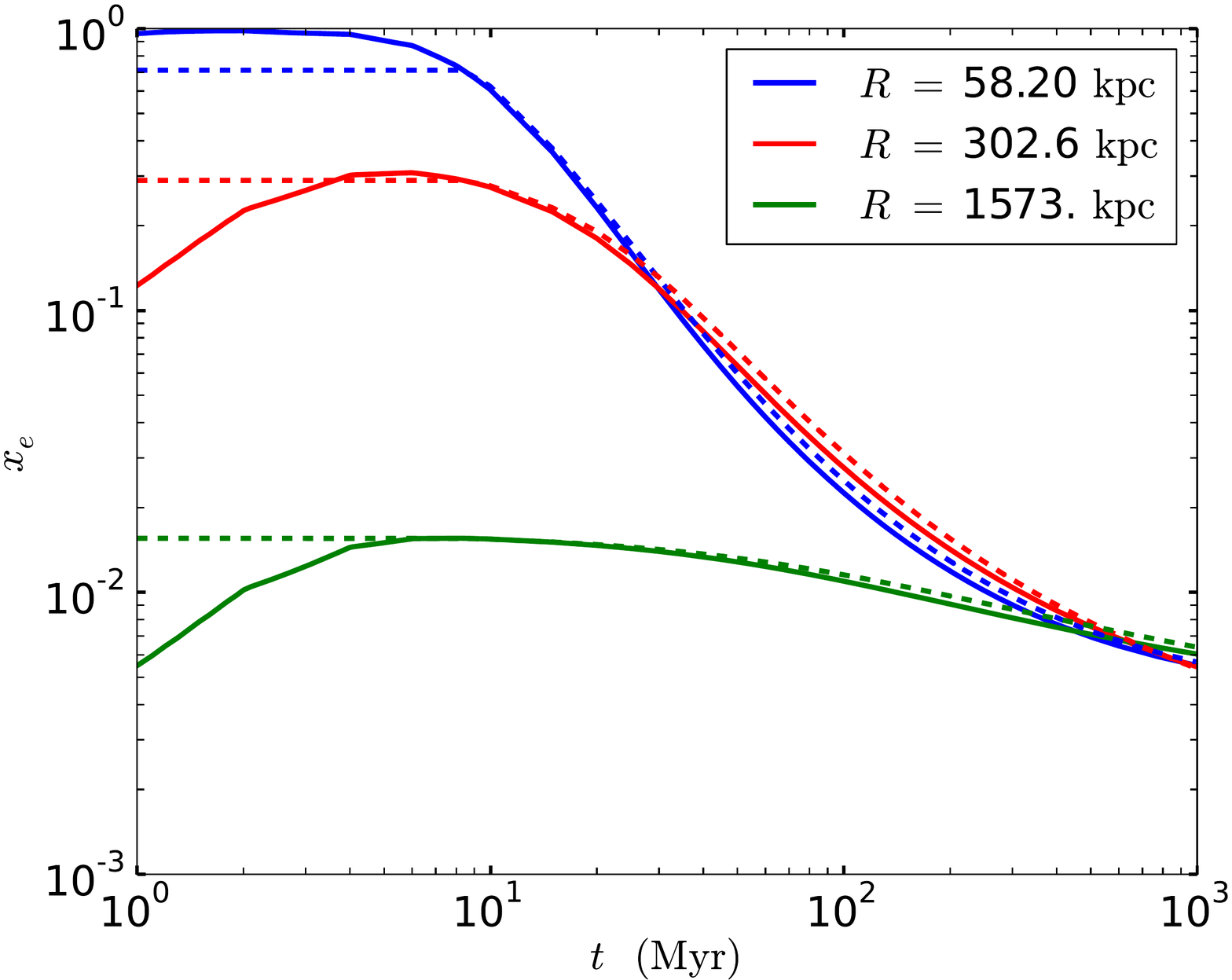}
\caption{(Left) An example of the simple analytical fit to the ionization profile outside a halo. The red lines represent the neutral and ionized fractions produced by the numerical simulation. The blue line represents the best fit of the analytical model. (Right) An example of the recombination model compared to the numerical simulation. The solid lines represent the electron fraction at four fixed radii as a function of time from the numerical simulations. The dashed lines represent the behavior of the electron fraction as described by~\eqref{eq:recom1}.}
\label{fig:fit}
\end{figure*}

A broad analysis of our suite of simulations showed that the ionization profile around a test halo during star formation was well fit by the following class of functions:
\begin{align}
x_e(R)&=\frac{1}{1+\left(\frac{R-A}{B}\right)^{1/C}}\label{eq:chiofr}\\
R(x_e)&=A+B\left(\frac{1}{x_e}-1\right)^{C}.\label{eq:rofchi}
\end{align}
The fit of Equation~\eqref{eq:chiofr} approximates the electron function as a function of radius for a fixed moment in time. The parameters $A$, $B$, and $C$ may be interpreted as a distance offset, a scale factor, and a power law, respectively (this sentence needs work). These parameters are, in principle, functions of the Hubble time,  time since star formation began, and luminosity of ionizing photons. An example of one such fit is given in Figure~\ref{fig:fit}(top). We have included the inverse of the model fit because our simulation calls for the radius at which the ionization reaches a given level, as given by Equation~\eqref{eq:rofchi}.

An analytic model for the time dependence of the scale radius $R_s$ of a Str{\"o}mgren sphere in an expanding universe is presented in \cite{Donahue:1987} (see also \cite{ShapiroG:1987}). Explicitly:
\begin{align}\label{eq:donahue}
R_s(t)&=\left(21\text{ kpc}\right)\left[\left(\frac{S_0}{10^{49}\text{ s}^{-1}}\right)\cdot f_1(t)\right]^{1/3}\\
f_1(t)&=\frac{(8.95\times10^5)\lambda}{(1+z_0)^3}\exp\left(\frac{\lambda}{t_c}\right)\nonumber\\
&\quad\quad\times\left[t_cE_2\left(\frac{\lambda}{t_c}\right)-E_2(\lambda)\right]\text{ Myr}\\
\Delta R_s(t)&=(3.2\text{ Mpc})\left(\frac{1+z}{10}\right)^{0.7}\nonumber\\
&\quad\quad\times\left(\frac{S_0}{10^{49}\text{ s}^{-1}}t\right)^{0.18}t_c^{-0.466}
\end{align}

Here, $\lambda = t_H/t_{\text{rec}}$, $t_c=1+t/t_{\text{rec}}$, and $E_n(x)$ is the exponential integral of the $n$th order. This more complicated model is needed to replace the simple Str{\"o}mgren sphere model because of the Hubble expansion freezes recombinations when the recombination time becomes longer than the Hubble time.

We found that the parameters $(A,B,C)$ of eq~\eqref{eq:rofchi} are well fit using the following formula:
\begin{align*}
A(t,S_0)&\propto S_0\cdot R_s(t)\\
B(t,S_0)&\propto S_0\cdot \Delta R_s(t)\\
C(t,S_0)&=2.0.
\end{align*}

\subsection{Ionization Profile After Star Formation}\label{sec:recomb}

\begin{figure*}
\includegraphics[width=8.7cm]{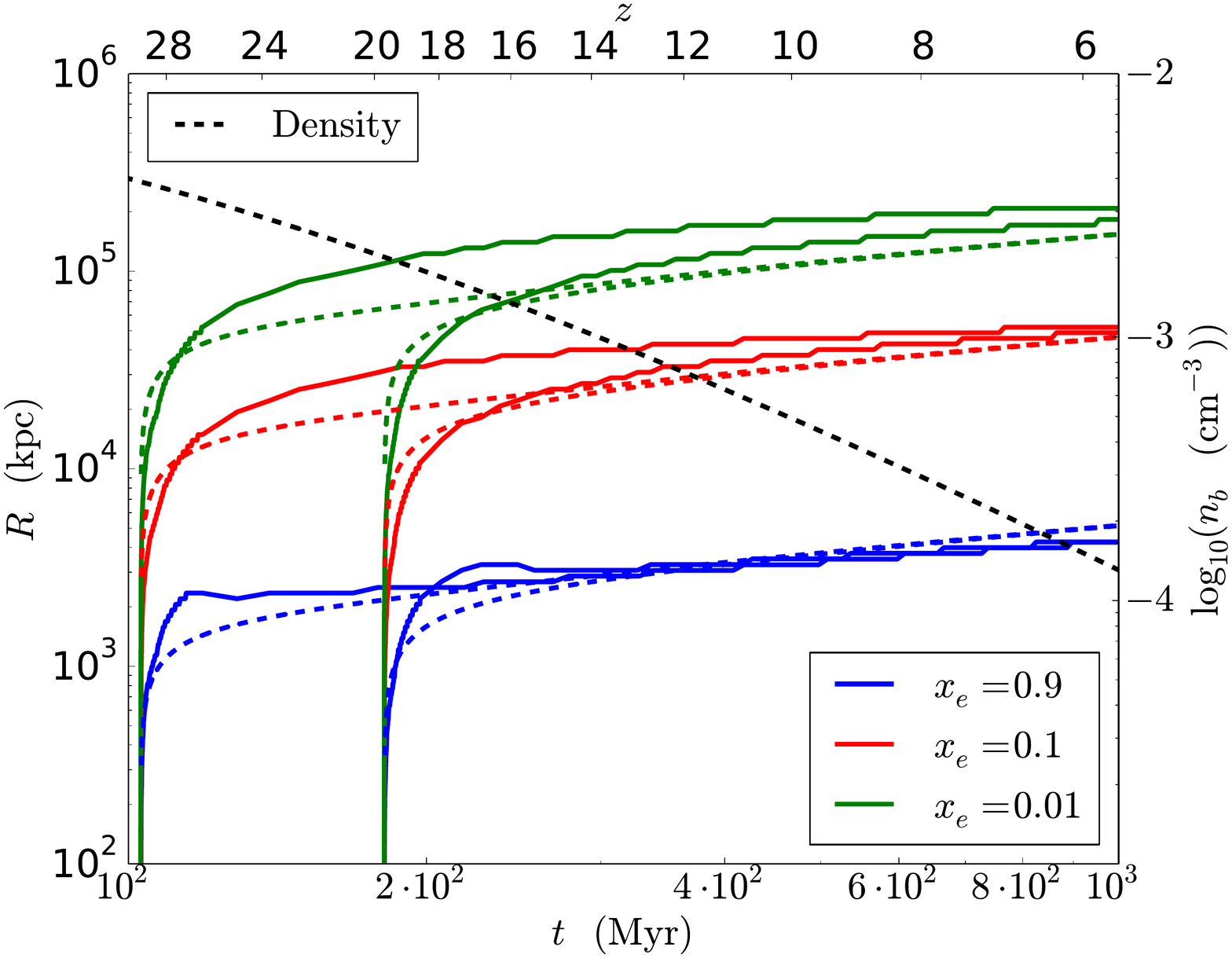}
\includegraphics[width=8.7cm]{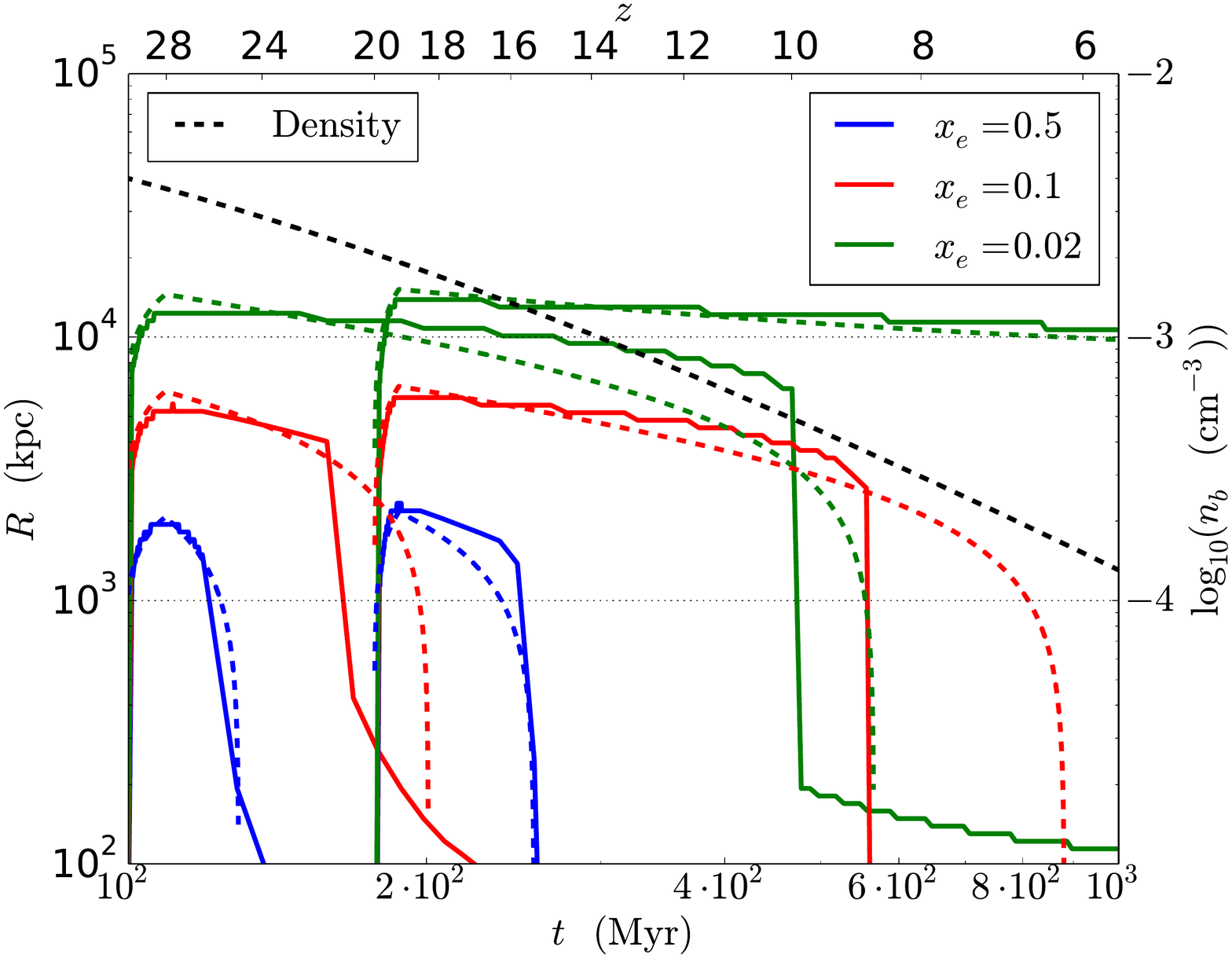}
\caption{Full analytic model for ionization profile for (left) continuous and (right) instantaneous SF against the numerical model for at initial redshifts $z_0=30,20$. The solid lines represent the radii at which a fixed electron fraction is found for the numerical model. The dotted lines represent the corresponding analytic fits. (left) For the continuous case, deviations observed are consistent with imperfections in the model as documented in \protect\cite{Donahue:1987}. 
(right) For the instantaneous case, deviations occur due to the spatial dependence of temperature.}
\label{fig:fullmodel}
\end{figure*}

When star formation ends, the ionized gas begins to recombine. This process is modeled by the differential equation $\dot{n_{\text{HI}}}=(\alpha n_e)n_{\text{HII}}$. Assuming a purely hydrogen medium, this equation is solved in an expanding universe by the following equation:
\begin{align}\label{eq:recom1}
x_e(R,t)&=\left(\int_{t_0}^{t}\alpha^{(2)} n(t)dt+\frac{1}{x_e(R,t_0)}\right)^{-1}.
\end{align}
 
The correction for this equation for the more general case of helium and hydrogen is well approximated by an multiplicative constant outside the integral that represents the extra electrons from ionized helium. The integral, in general, is impossible to solve without an analytic expression for the temperature. However, the recombination coefficient $\alpha$ depends weakly on temperature ($\alpha\propto T^{-0.7}$), we may assume $\alpha$ is constant as an approximation. The integral may then be solved explicitly:
\begin{align}\label{eq:recom2}
\int_{t_0}^{t} n(t')dt'=\frac{2n_{0}\Omega_b}{3H_0\Omega_m^{1/2}}\left[(1+z_{0})^{3/2}-(1+z)^{3/2}\right].
\end{align}

This formula provides an excellent description of the time domain behavior of the electron fraction. A comparison between this model and a radiative transfer simulation is shown in Figure~\ref{fig:fit}(bottom).

In the case of continuous star formation, the initial profile expands according to the time dependent scale in \eq~\eqref{eq:donahue}. An example of the resulting fit is plotted in Figure~\ref{fig:fullmodel} against two examples of continuous star formation produced from the suite of numerical simulations. The analytic model captures both the transient behavior as the sphere initial expands and the long term steady expansion of sphere as recombinations freeze out with the expansion of the universe.

In the case of instantaneous star formation, the initial profile expands according to \eq~\eqref{eq:donahue} while the brightest stars remain alive. After the luminosity begins to plummet as massive stars die, the ionized region begins decaying according to \eq~\eqref{eq:recom1} and \eq~\eqref{eq:recom2}. An example of the resulting fit is plotted in Figure~\ref{fig:fullmodel} against two examples of instantaneous star formation produced from the suite of numerical simulations. The analytic model captures both the growth and decay phases of the region. The largest deviation is the time taken by the outer regions to decay as compared to the numerical model; this is due to the dependence of temperature on radius.

\subsection{Time-dependent Spectral Energy Distribution }\label{sec:sed}

A numerical model for the evolution of the ionization in a region around a source of ionization is presented in \cite{Ricotti:2001}. This numerical model takes a spectral energy distribution (SED) and outputs the density profile of neutral and ionized hydrogen (as well as several other elements) as a function of time.

In our statistical simulation, we take a star forming halo to be a point source of ionization. We make use of the code from \cite{Ricotti:2001} to simulate the behavior of the IGM around a model point source for a suite of $(S_0,z_0)$ values, where $S_0$ is the rate of ionizing photon production and $z_0$ is the redshift of initial star formation. The SED used for these simulations are produced using the Starburst99 code \citep{Leitherer:1999} with the lowest possible metallicity ($Z=0.001$). 

The suite of $(S_0,z_0)$ values was chosen to be large enough to cover the range of relevant luminosities and redshifts in our statistical simulation. The number of values in each range was increased until the change in behavior between adjacent luminosity and redshift points changed continuously and predictably.

\subsection{Simulation Details}

The simulation is divided into 72 intervals with $\Delta T = 12.5$ Myr. The start of each time interval is denoted $T_i$. At $T_0$, a number $k$ of halos with mass $M_{\text{added}}$ are added following a Poisson distribution with average masses extracted from a Sheth-Tormen distribution \citep{PressS:1974,ShethT:2002}:
\begin{align}
f(k;\lambda)&=P(M<M_{\text{added}}<M+dM)=\frac{\lambda^ke^{-\lambda}}{k!}\\
\lambda dM&=N(M_{\text{added}},T_0)dM\quad\text{(Sheth-Tormen)}.
\end{align}
For later times, halo counts are drawn by the same Poisson process with $\lambda = N(M_{\text{added}},T_i)-N(M_{\text{added}},T_{i-1})$. Thus, at any time $T_i$, the total population is drawn from Poisson distribution with $\lambda=N(M_{\text{added}},T_i)$, consistent with the Sheth-Tormen model.

\subsection{Convergence Tests}

The statistical simulation requires the discretization of several continuous variables. We decreased the spacing of these variables systematically until convergence of the simulation results was observed. Here we discuss in depth the relevant details of this process for all of the relevant variables:
\begin{itemize}
\item Electron fraction spacing ($\Delta x$): The minimum electron fraction $x_0$ was chosen to be $0.001$. At this electron fraction, $\Delta\tau_e$ between $Q=1$ and $Q=0$ integrated across the redshift range of our simulation is negligible, so that smaller electron fractions are irrelevant. The maximum electron fraction $x_{m-1}$ was chosen to be $0.9$. Our analytic model for the profile produces $r(x)$, where $r\rightarrow0$ as $x\rightarrow0$. The radius at which $x=0.9$ in the analytic model is a good representation of the area of total ionization within the simulation. The number of $x_i$ points was increased until the results were found to converge beyond $m=20$.

\item Halo addition time step ($\Delta T$): Halos are added to the simulation at constant time intervals. It was found that the results converged beyond $50$ intervals, but computational complexity does not increase significantly with the number of time intervals, so a value of $72$ was settled on.

\item Output time step ($\Delta t$): The time interval between output reports does not have any effect on the results of the simulation, so convergence is not relevant.

\item Comoving volume ($V$): Computational complexity scales linearly with the volume of the box. We found that a volume of $V=10^6\text{ Mpc}^3$ produced sufficiently large halos to ensure complete reionization by a redshift of $z\sim 5.8$. Larger simulations are possible, but require significant time resources and will be performed moving forward.
\end{itemize}

\label{lastpage}
\end{document}